\documentclass{aa}
\usepackage{psfig}
\usepackage{natbib}
\bibpunct{(}{)}{,}{a}{}{,} 
\usepackage{graphicx}

\def\la{\;
\raise0.3ex\hbox{$<$\kern-0.75em\raise-1.1ex\hbox{$\sim$}}\; }
\def\ga{\;
\raise0.3ex\hbox{$>$\kern-0.75em\raise-1.1ex\hbox{$\sim$}}\; }

\newcommand{\dmm}{$\Delta\mu/\mu\,$}
\newcommand{\daa}{$\Delta\alpha/\alpha\,$}
\newcommand{\kms}{km~s$^{-1}\,$}

\newcommand{\ms}{m~s$^{-1}\,$}

\newcommand{\cmm}{cm$^{-3}\,$}
\newcommand{\etal}{{et al.}}
\newcommand{\amm}{NH$_3$}
\newcommand{\nnhp}{N$_2$H$^+$}
\newcommand{\hcccn}{HC$_3$N}

\begin{document}

\title{
Searching for chameleon-like scalar fields with the ammonia method\thanks{Based 
on observations obtained with the Medicina 32-m telescope operated by
INAF~-- Istituto di Radioastronomia, the 100-m telescope of the Max-Planck
Institut f\"ur Radioastronomie at Effelsberg, and the Nobeyama Radio
Observatory 45-m telescope of the National Astronomical Observatory
of Japan. 
} 
}
\subtitle{}
\author{S. A. Levshakov\inst{1,2}
\and
P. Molaro\inst{1}
\and
A. V. Lapinov\inst{3}
\and
D. Reimers\inst{4}
\and
C. Henkel\inst{5}
\and
T. Sakai\inst{6}
}
\institute{
INAF-Osservatorio Astronomico di Trieste, Via G. B. Tiepolo 11,
34131 Trieste, Italy
\and
Ioffe Physical-Technical Institute, 
Polytekhnicheskaya Str. 26, 194021 St.~Petersburg, Russia\\
\email{lev@astro.ioffe.rssi.ru} 
\and
Institute for Applied Physics, Uljanov Str. 46, 603950 Nizhny Novgorod, Russia
\and
Hamburger Sternwarte, Universit\"at Hamburg,
Gojenbergsweg 112, D-21029 Hamburg, Germany
\and
Max-Planck-Institut f\"ur Radioastronomie, Auf dem H\"ugel 69, D-53121 Bonn, Germany
\and
Institute of Astronomy, The University of Tokyo, Osawa, Mitaka, Tokyo  
181-0015, Japan
}
\date{Received 00  ; Accepted 00}
\abstract
{}
{
We probe the dependence of the electron-to-proton mass ratio, $\mu = m_{\rm e}/m_{\rm p}$,
on the ambient matter density by means of radio astronomical observations.
} 
{The ammonia method, which has been proposed to explore the electron-to-proton mass ratio,   
is applied to nearby dark clouds in the Milky Way.
This ratio, which is measured in different physical environments of 
high (terrestrial) and low (interstellar) densities of baryonic matter is 
supposed to vary in chameleon-like scalar field models, which predict
strong dependences of both masses and coupling constant on the local matter density. 
High resolution spectral observations of molecular cores
in lines of NH$_3$ $(J,K) = (1,1)$,  
HC$_3$N $J = 2-1$, and N$_2$H$^+$ $J = 1-0$ were performed at three
radio telescopes to measure the radial velocity offsets,
$\Delta V \equiv V_{\rm rot} - V_{\rm inv}$,
between the inversion transition of NH$_3$ (1,1) and the rotational transitions
of other molecules with different sensitivities
to the parameter \dmm $\equiv (\mu_{\rm obs} - \mu_{\rm lab})/\mu_{\rm lab}$.
}
{The measured values of $\Delta V$ exhibit a statistically significant velocity
offset of $23\pm4_{\rm stat}\pm3_{\rm sys}$ \ms. When interpreted in terms
of the electron-to-proton mass ratio variation, this infers that
\dmm = $(2.2\pm0.4_{\rm stat}\pm0.3_{\rm sys})\times10^{-8}$.
If only a conservative upper bound is considered, then the maximum 
offset between ammonia and the other molecules is
$|\Delta V| \leq 30$ \ms.
This provides the most  accurate reference point at $z = 0$ for \dmm\ of 
$|\Delta \mu/\mu| \leq 3\times10^{-8}$. 
}
{}
\keywords{Line: profiles -- ISM: molecules -- Radio lines: ISM -- Techniques:
radial velocities}
\authorrunning{S. A. Levshakov \etal}
\titlerunning{Searching for chameleon-like scalar fields with the ammonia method} 

\maketitle

\section{Introduction}
\label{sect-1}

Testing the variability of dimensionless physical constants 
is an important topic in contemporary laboratory and astrophysical experiments.
The physical constants are not supposed to vary in the Standard Model (SM) of
particle physics but do vary quite naturally in grand unification theories,
multidimensional theories, and whenever there is a coupling between light
scalar fields and baryonic matter. In particular, light scalar fields 
have been widely discussed in the context of
dark energy, since they provide negative pressure that may be responsible
for the cosmic acceleration detected at low redshifts, $z<1$
(Caldwell \etal\, 1998; Peebles \&  Rata 2003).
If this scalar field does exist, then a question arises: why has it 
not been detected in local
tests of the equivalence principle or fifth force searches? A solution was suggested 
using the so-called `chameleon' models (Khoury \& Weltman 2004; Brax \etal\ 2004;
Mota \& Shaw 2007). 
These models assume that a light scalar field acquires both an effective potential 
and effective mass because of its coupling to matter that strongly depends on the 
ambient matter density. In this way, this 
scalar field may evade local tests of the equivalence principle and fifth force experiments
since the range of the scalar-mediated fifth force for the terrestrial 
matter densities is too small
to be detected.
This is not the case for space-based tests, where the matter density is considerably lower,
an effective mass of the scalar field is negligible,
and an effective range for the scalar-mediated force is large.
The present paper deals with one such astronomical test based on spectral observations
of molecular clouds in the Milky Way disk. 
Additional tests employing polarization of the light from the stars and a modification of the
Sunyaev-Zel'dovich effect in the cosmic microwave background 
due to a coupling between a chameleon-like scalar field and photons
are described, respectively, in Burrage \etal\ (2009) and Davis \etal\ (2009). 

Astronomical spectroscopy can probe the physical constants that describe
atomic and molecular discrete spectra: 
the fine-structure constant, $\alpha = e^2/(\hbar c)$, and the electron-to-proton mass ratio,
$\mu = m_{\rm e}/m_{\rm p}$. 
In GUTs, these constants mediate the strength of fundamental
forces: $\alpha$ is the coupling constant of the electromagnetic interaction,
$m_{\rm e}$ is related to the vacuum expectation value of the Higgs field, namely
to the scale of the weak nuclear force, and $m_{\rm p}$ is proportional to the
quantum chromodynamics scale.
We note that the predicted variabilities of the fine-structure constant
and the electron-to-proton mass ratio
are not independent and that the variations in $\mu$ may exceed those of $\alpha$
(Calmet \& Fritzsch 2002; Langacker \etal\ 2002; Dine \etal\ 2003; 
Flambaum \etal\ 2004). 
Thus, a hypothetical variation in $\mu$ is expected to be easier to detect than that in $\alpha$.

Despite many efforts,
the variability in $\alpha$ and $\mu$ has never been detected.
Data obtained from high precision frequency measurements in laboratory
experiments with atomic clocks and from  astronomical observations provide only upper limits.
For example, laboratory experiments have delivered the following results:
fractional temporal variations in $\mu$ are restricted to a level of
$\dot{\mu}/\mu = (3.8\pm5.6)\times10^{-14}$ yr$^{-1}$ (Shelkovnikov \etal\, 2008), and 
$\dot{\mu}/\mu = (1.6\pm1.7)\times10^{-15}$ yr$^{-1}$ (Blatt \etal\, 2008),  
whereas the current level for $\alpha$ is
$\dot{\alpha}/\alpha = (-1.6\pm2.3)\times10^{-17}$ yr$^{-1}$ (Rosenband \etal\, 2008).
Here $\Delta \mu$ and $\Delta \alpha$ are the relative changes between the values
measured at two different epochs. 

The most stringent upper limits to \daa\, and \dmm\,
obtained from astronomical observations of extragalactic objects 
are restricted by a few units of ppm (1ppm = $10^{-6}$). 
Here $\Delta \alpha/\alpha = (\alpha' - \alpha)/\alpha$, where $\alpha, \alpha'$
denote the values of the fine-structure constant in the laboratory and the
specific absorption/emission line system of a Galactic or extragalactic object
(the same definition is applied to \dmm).
However, there are claims of a variability in both $\alpha$ and $\mu$ at 
the 5$\sigma$ and 4$\sigma$ confidence levels, respectively, although they are in contrast 
to other null results and the whole issue is highly controversial.
The current observational status is the following.

From the measurements of the relative radial velocity shifts of different 
absorption lines (e.g., \ion{Mg}{ii}, \ion{Si}{ii}, \ion{Fe}{ii}, \ion{Zn}{ii}) of 
143 QSO absorption systems obtained with the Keck/HIRES in the redshift range
$0.2 < z < 4.2$, 
Murphy \etal\ (2004) claimed that $\Delta\alpha/\alpha = -5.7\pm1.1$ ppm. 
On the other hand, Chand \etal\ (2004)
using VLT/UVES spectra of bright QSOs analyzed 23 absorption systems, which are not
in common with those studied by the Keck telescope, and failed to reproduce Murphy \etal's result. 
At first, they claimed a very stringent limit of 
$\Delta\alpha/\alpha = -0.6\pm0.6$ ppm, 
but Murphy \etal\ (2008a) demonstrated convincingly that the error estimate of
Chand \etal\ (2004) is underestimated. 
When these errors are properly accounted for, the new weighted mean becomes 
$\Delta\alpha/\alpha = -6.4\pm3.6$ ppm. 
The new value has a scatter that is larger than the quoted errors implying 
that there are systematic errors that are comparable or even larger
than the statistical ones. 
However, Srianand \etal\ (2008) showed that excluding two systems that deviate
from the mean at the 3$\sigma$ level when reanalyzing VLT/UVES systems
leads to $\Delta\alpha/\alpha = 0.1\pm1.5$ ppm. 
We note that the only claimed non-zero measurement of \daa\ corresponds to an uncertainty
in the wavelength scale calibration smaller 100 \ms. This precise
calibration of the Keck/HIRES spectra is difficult to achieve as 
shown by Griest \etal\ (2010). 
Using iodine exposures to calibrate the normal Th-Ar Keck data pipeline output, 
they found absolute wavelength offsets of 500 \ms to 1000 \ms with drifts of 
more than 500 \ms over a single night, 
and drifts of nearly 2000 \ms over several nights. Their  
conclusion is that these systematic uncertainties make it difficult to 
use Keck/HIRES QSO spectra 
to constrain the change in the fine structure constant at the $10^{-6}$ level.
The most stringent constraint on $|\Delta\alpha/\alpha| < 2$ ppm 
was found from the \ion{Fe}{ii} lines at $z = 1.15$ towards the bright quasar 
HE~0515--4414 (Quast \etal\ 2004; Levshakov \etal\ 2005; Molaro \etal\ 2008). 

Bounds on  $\mu$ variations are most effectively obtained from observations of the Werner and Lyman
series of the molecular hydrogen H$_2$ in damped Ly-$\alpha$ systems (DLA). 
The electron-vibro-rotational transitions have
different dependences on the reduced mass and can be used to constrain the variability 
of $\mu$ (Thompson 1975; Varshalovich \& Levshakov 1993).
The measurements of $\mu$ rely on the same H$_2$ systems observed  with
VLT/UVES and claim a variability of $\Delta\mu/\mu = 24\pm6$ ppm (Reinhold \etal\ 2006)
or no variability with 
$-15\ {\rm ppm} < \Delta\mu/\mu < 57$ ppm (Levshakov \etal\ 2002),
$|\Delta\mu/\mu| \leq 49$ ppm (Wendt \& Reimers 2008), 
$\Delta\mu/\mu = 2.6\pm3.0$ ppm (King \etal\ 2008), and
$\Delta\mu/\mu = -7\pm8$ ppm (Thompson \etal\ 2009).
Radio astronomical observations place constraints on $\mu$-variations
of $|\Delta \mu/\mu| < 1.8$ ppm at redshift $z = 0.68$ (Murphy \etal\ 2008b),
and $|\Delta \mu/\mu| < 0.6$ ppm at $z = 0.89$ (Henkel \etal\ 2009).  

One should keep in mind,
however, that laboratory experiments and spectra of extragalactic objects probe
very different time-scales and different regions of the universe,
and the connection between them is quite model dependent
(Mota \& Barrow 2004).

Astronomical measurements of the dimensionless constants are based on the comparison
of the line centers in the absorption/emission spectra of astronomical objects
and the corresponding laboratory values. 
To distinguish the line shifts due to the radial  
motion of the object from those caused by
the variability of constants, lines with different
sensitivity coefficients, $Q$,  to the variations in $\mu$ and/or $\alpha$ should be employed. 
Unfortunately, the observable optical lines have very close
sensitivity coefficients with differences of $|\Delta Q|$ that do not exceed 0.05.
Combined with the resolving power of spectrographs at modern optical telescopes, 
this small difference leads to an upper limit on \daa\, and \dmm\, of $\sim$1 ppm,  
the optimal value  
achievable in observations of extragalactic objects with existing facilities.
However, one can probe the values of fundamental constants at a considerably more
accurate level if, firstly, nearby objects in the Milky Way are observed and, secondly,
spectral lines from other frequency ranges, infrared and radio, are employed.
This statement is based on the following considerations.

As mentioned above, the scalar field models suggest
a coupling between the scalar fields and baryonic matter. This coupling
results in a functional dependence of $\mu$ and $\alpha$ on $\rho$, 
the local matter density (Olive \& Pospelov 2008).
This coupling alone may prevent a positive detection of the variations in $\mu$ and
other dimensionless constants in laboratory studies, since they are performed under the
same terrestrial conditions 
and the effective range of the scalar-field-mediated force is smaller than 1 mm at
terrestrial matter densities (Olive \& Pospelov 2008). 
On the other hand, the density in cold molecular
clouds is only  $10^3 - 10^5$ particles per cm$^3$, i.e., $\sim 10^{19}$ times lower than 
in terrestrial environments.
Because of this extremely large difference between the matter
densities non-zero values of \dmm\ and \daa\ are predicted. 
It is very important to compare the above ratio of $\sim10^{19}$ 
with the differences in $\rho$
between molecular clouds themselves, to find that they are negligible, i.e., 
at most one or two orders of magnitude for a given tracer. 
In the context of \dmm\ (or \daa) this means that
measurements of all dense molecular clouds are identical irrespective of their location
in space and time (redshift). 
Thus, nearby Galactic molecular clouds can be observed to ensure 
a strong signal, i.e., that the line profiles can be centered with high accuracy.

Molecular lines originating in these clouds are mainly observed in cm and mm radio bands.
An advantage of radio observations is that 
very narrow spectral lines (of line widths $\sim$100 \ms) arising 
in cold molecular cores can be observed with
extremely high spectral resolution, FWHM $\sim25$ \ms.  
In optical observations of extragalactic objects, 
we have in general broader sources, $b \ga 1$ \kms,
and use lower resolution spectrographs, FWHM $\ga 4$ \kms\ (Levshakov \etal\ 2007).  
Taking into account that the uncertainty
in the line position is roughly 1/10th of the pixel size, this infers a gain of
about two orders of magnitude in the accuracy of the measurements of \daa\ and/or \dmm\ if
radio spectra of the local interstellar objects are used.

Complementary to this, differences in the sensitivity coefficients for lines from the  microwave,
far IR, and radio ranges are much larger than those from optical and UV ranges.
For example, the difference between the sensitivity coefficient of the 
$(J,K)=(1,1)$ inversion transition 
of ammonia (\amm) and that of any rotational transition in another molecule is
$\Delta Q = 4.46 - 1 = 3.46$ (Flambaum \& Kozlov 2007).
When compared with UV transitions, this infers a gain of about 70 times in the sensitivity of the
line positions to the change in $\mu$.

Our preliminary study of cold molecular clouds (Levshakov \etal\  2008a, hereafter LMK)
was based on published results of high quality spectral radio observations obtained
with the 100-m Green Bank telescope (Rosolowsky \etal\ 2008; Rathborne \etal\ 2008)
and the 45-m Nobeyama radio telescope (Sakai \etal\ 2008).
Comparison of the relative radial velocities of the \amm\ $(J,K) = (1,1)$ inversion
transition and CCS $J_N = 2_1 - 1_0$ rotational transition measured a systematic shift of 
$\Delta V \equiv V_{lsr}({\rm CCS}) - V_{lsr}({\rm NH}_3) \sim 60$ \ms,
which when interpreted in terms of the electron-to-proton mass ratio
variation infers that \dmm $\sim 6\times10^{-8}$.
We also noted that a similar offset between \nnhp\ and \amm\ measured by
Pagani \etal\ (2009) in the cold dark cloud L183 could have a similar physical
origin (Molaro \etal\ 2009). 

In this paper, we present results of our own spectral observations 
of cold and compact molecular cores in the
Taurus giant molecular complex obtained with
the Medicina 32-m telescope, the Effelsberg 100-m telescope, and
the Nobeyama 45-m telescope.

\section{The ammonia method}
\label{sect-2}

Narrow molecular lines observed in cold dark clouds provide 
a sensitive spectroscopic tool to study relative 
shifts of the order of a few 10 \ms\
between radial velocities of different molecular transitions.
Among numerous molecules detected in the interstellar medium,
ammonia (NH$_3$) is of particular interest because of the high
sensitivity of the inversion frequency to a change in $\mu$.
The inversion vibrational mode of NH$_3$ is described by a double-well
potential, the first two vibrational levels lying below the barrier.
The quantum mechanical tunneling splits these two levels into inversion doublets
providing a transition frequency that falls in the microwave range
(Ho \& Townes 1983).

For the ammonia isotopologue $^{15}$ND$_3$, 
van Veldhoven \etal\ (2004) first showed 
that the inversion frequency of the $(J,K) = (1,1)$ level varies as\footnote{The sign of the
sensitivity coefficient ${\cal Q} = 5.6$ was misprinted in van Veldhoven \etal\ (2004),
as noted by Flambaum \& Kozlov (2007).} 
\begin{equation}
\left( \frac{\Delta \nu}{\nu} \right)_{\rm inv}
= 5.6\frac{\Delta \mu}{\mu}\ ,
\label{am0}
\end{equation}
i.e., the inversion transition is an order of magnitude more sensitive to $\mu$-variation than
molecular vibrational transitions, which scale as $E_{\rm vib} \sim \mu^{1/2}$. 

The sensitivity coefficient of
the inversion transition $(J,K) = (1,1)$ in \amm\ ($\nu = 23.7$ GHz) 
was calculated by Flambaum \& Kozlov (2007)
from the numerical integration of the Schr\"odinger equation for different
values of $\mu$, and from the analytical
Wentzel-Kramers-Brillouin (WKB) approximation of the
inversion frequency. Both methods provide similar results giving 
\begin{equation}
\left( \frac{\Delta \nu}{\nu} \right)_{\rm inv}
\equiv \frac{\tilde{\nu} - \nu}{\nu}
= 4.46\frac{\Delta \mu}{\mu}\ .
\label{am1}
\end{equation}
Here $\nu$ and $\tilde{\nu}$ are the frequencies corresponding to the
laboratory value of $\mu$ and to an altered  $\mu$ in a low-density
environment, respectively. 

The rotational frequency scales as $E_{\rm rot} \sim \mu$ and, thus,
\begin{equation}
\left( \frac{\Delta \nu}{\nu} \right)_{\rm rot}
\equiv \frac{\tilde{\nu} - \nu}{\nu}
= \frac{\Delta \mu}{\mu}\ .
\label{am2}
\end{equation}
By comparing the observed inversion frequency of NH$_3$ (1,1) with a suitable
rotational frequency of another molecule produced {\it co-spatially} with ammonia,
a limit on the spatial variation of $\mu$ can be determined. 

In radio astronomical observations, any frequency shift $\Delta \nu$
is related to the radial velocity shift $\Delta V_r$
($V_r$ is the line-of-sight projection of the velocity vector)
\begin{equation}
\frac{\Delta V_r}{c} \equiv \frac{V_r - V_0}{c}
= \frac{\nu_{\rm lab} - \nu_{\rm obs}}{\nu_{\rm lab}}\ ,
\label{am3}
\end{equation}
where $V_0$ is the reference radial velocity, and
$\nu_{\rm lab}$, $\nu_{\rm obs}$ are 
the laboratory and observed frequencies, respectively. Therefore, by comparing the
apparent radial velocity $V_{\rm inv}$ for the NH$_3$ inversion transition
with the apparent radial velocity  $V_{\rm rot}$ for rotational lines
originating in the same molecular cloud and moving with
a radial velocity $V_0$ with respect to the local
standard of rest, we can
find from Eqs.~(\ref{am1} - \ref{am3}) 
\begin{equation}
\frac{\Delta \mu}{\mu} = 0.289\frac{V_{\rm rot} - V_{\rm inv}}{c}
\equiv 0.289\frac{\Delta V}{c}\ .
\label{am4}
\end{equation}

From Eq.(\ref{am4}), we can estimate the limiting accuracy of \dmm\ available
from modern radio astronomical observations.
At high spectral resolution in the microwave range (FWHM $\sim 30$ \ms),
the errors in the molecular line position measurements
are mainly restricted by the uncertainties
in laboratory frequencies, $\varepsilon_\nu \sim 0.1-1$ kHz,
which correspond to the $V_{lsr}$ uncertainties of
$\varepsilon_v \sim 1-10$ \ms.
Taking into account that \dmm $\sim 0.3\Delta V/c$, the expected accuracy of \dmm\ 
in a single measurement is about $10^{-8}$. 
This level of accuracy
can be achieved, however, only under ideal conditions since it assumes that
molecules are identically
distributed within the cloud and are observed
simultaneously with the same receiver, beam size, system temperature, and
velocity resolution.
Violation of any of these conditions leads to random
shifts of the line centers, which are
referred to as the {\it Doppler noise}.
The input of the Doppler noise to a putative \dmm\ signal can
be reduced to some extent if the velocity shifts due to
the inhomogeneous distribution of molecules and to
instrumental imperfections are minimized.
For our purposes, we express
the velocity offset $\Delta V$  in Eq.(\ref{am4}) as the
sum of two components 
\begin{equation}
\Delta V = \Delta V_\mu + \Delta V_n, 
\label{am5}
\end{equation}
where $\Delta V_\mu$ is the shift due to $\mu$-variation and $\Delta V_n$
is the Doppler noise.
We assume that the noise component has zero mean
$\langle \Delta V_n \rangle = 0$ \kms and a finite variance.
The signal $\Delta V_\mu$ can then be estimated statistically by
averaging over a data sample
\begin{equation}
\langle \Delta V \rangle = \langle \Delta V_\mu \rangle,\\ 
Var(\Delta V) = Var(\Delta V_\mu) + Var(\Delta V_n).
\label{am6}
\end{equation}
From this equation it is seen that the measurability of the signal ($\Delta V_\mu$)
depends critically on the value of the Doppler noise.
The Doppler noise $Var(\Delta V_n)$ 
can be reduced by an appropriate choice of molecular lines and by
special criteria applied to the selection of targets.

\subsection{Molecules appropriate for \dmm\ measurements}
\label{sect-2-1}

The observed molecular transitions should 
as far as possible share the same volume elements
to have similar Doppler velocity shifts.
The ammonia inversion transitions are usually detected
in dense molecular cores ($n \ga 10^4$ \cmm), which are represented in the
Milky Way disk by a large variety of types and physical
properties (Di Francesco \etal\ 2007).
Mapping of the dense molecular cores in different molecular lines shows that
there is a good correlation between the ammonia NH$_3$,
\nnhp, and \hcccn\ distributions   
(e.g., Fuller \& Myers 1993; Hotzel \etal\ 2004; 
Tafalla \etal\ 2004; Pagani \etal\ 2009).
However, in some clouds \amm\ is not traced by \hcccn.
The most striking case is the  
dark cloud TMC-1, where peaks of line emission are offset
by 7 arcmin (Olano \etal\ 1988). 

In general,
N-bearing molecules such as \amm\ and \nnhp\ 
trace the inner cores, where the density approaches $10^5$ \cmm. At  
the same time, the carbon-chain molecules disappear from the gas-phase
because of freeze-out onto dust grains (Tafalla \etal\ 2004).
HC$_3$N, as well as other C-bearing molecules, 
are usually distributed in the outer parts of the cores. 
The mutual distribution of \amm\ and \hcccn\ is affected
by chemical differentiation in the process of the dynamical evolution
of the core.
HC$_3$N is abundant in the early evolutional stage
of the star-forming regions (Lee \etal\ 1996), 
when the fractional abundance of HC$_3$N remains almost
constant 
and the spatial distributions of the N- and C-bearing
molecules match each other quite well (Suzuki \etal\ 1992).
As the gas density increases and at dust temperatures $T_{\rm d} < 20$ K,
these distributions diverge
because of adsorption of the heavy molecules (e.g., \hcccn) from the gas phase
onto grain mantles (Flower \etal\ 2006).
In the later stages of the protostellar collapse, carbon
chain molecules are destroyed
by high velocity outflows and radiation from protostars, whereas the same
processes favor the desorption of ammonia from dust grains (Suzuki \etal\ 1992).

The chemical differentiation and velocity gradients within the molecular core 
are the main sources of the unavoidable Doppler noise in Eq.(\ref{am5}). 
Additional scatter in the $\Delta V_n$ values 
is caused by the 
different optical depths of the hyperfine structure transitions. 
However, all of these effects lead to the radial velocity shifts detected
between \amm\ and other molecules (\hcccn, \nnhp), which may be random from core to
core. Thus, being averaged over a large sample of targets, the Doppler noise
component $\Delta V_n$ in Eq.~(\ref{am5}) should be canceled out.

\section{Observations}
\label{sect-3}

For relative velocity measurements, the molecular lines
of \amm, \hcccn, and \nnhp\ were chosen. 
From the published data,
we selected 41 molecular cores with compact morphology,
i.e., the cores whose 
geometrical structure can be represented by a central
region of nearly constant density and a surrounding envelope
with a density that decreases as a power law.
For the dark clouds in the Taurus molecular complex,
a typical size of the central core is $r_0 \sim 20''-60''$ 
(e.g., Daniel \etal\ 2007). 
The selection was based on narrow and sufficiently strong
molecular lines that correspond to
individual hyperfine transitions:
\amm\ $(J,K) = (1,1)$, \hcccn\ $J=2-1$, and \nnhp\ $J=1-0$. 
This suggests that the selected cores have
low kinetic temperatures ($T_{\rm kin} \sim 10$ K) and 
are located at distances below 250 pc. 
The kinetic temperature is known to be surprisingly uniform in dark
clouds (e.g., Dickman 1975; Walmsley \& Ungerechts 1983). 
The list of sources is given in Table~\ref{tbl-1}.
Observations were performed with radio telescopes in Medicina,
Effelsberg, and Nobeyama between November 2008
and April 2009.

{\it Medicina.}\, Observations at the Medicina 32-m telescope\footnote{The 32-m VLBI antenna at 
Medicina is operated by the INAF-Istituto di Radioastronomia in Bologna.}
were carried out on November 24-28, 2008.
Both available digital spectrometers ARCOS (ARcetri COrrelation Spectrometer)
and MSpec0 (high resolution digital spectrometer) were used
with channel separations of 4.88~kHz and 2~kHz, respectively.
For ARCOS, this corresponds to 62 \ms\ 
at the position of the ammonia inversion transition (23 GHz) and 80 \ms\
at the rotational \hcccn\ (2--1) line (18 GHz). For MSpec0, it is
25 \ms\  and 32 \ms\ at the corresponding frequencies.
Only a few brightest objects (marked by a symbol $m$ in Table~\ref{tbl-1}) 
were observed. 
The Medicina 32-m telescope angular resolution is $\sim$1.6$'$ at 23 GHz
and $\sim$2.1$'$ at 18 GHz.
The pointing accuracy was superior to $25''$.
Spectra were taken in a position switching mode with a typical integration time
of 5 min for both ON- and OFF-source scans.
The OFF position was taken to be approximately five beam widths to the west 
of the source position. Typically, 10-20 ON/OFF pairs were taken,
depending on the frequency and source flux.
Unfortunately, because of poor weather conditions not all 
observations studied both \amm\ and \hcccn\ transitions.
Standard data reduction was performed using the CLASS reduction
package\footnote{http://www.iram.fr/IRAMFR/GILDAS/}.

{\it Effelsberg.}\, The ($J$,$K$) = (1,1) inversion line of ammonia (NH$_3$) and the
$J$=2--1 rotation line of cyanoacetylene (HC$_3$N)
were also observed with the 100-m telescope
at Effelsberg\footnote{The 100-m telescope at Effelsberg/Germany is operated
by the Max-Planck-Institut f{\"u}r Radioastronomie on behalf of the
Max-Planck-Gesellschaft (MPG).} on February 20-22, 2009. 
The corresponding targets are marked by symbol $e$ in Table~\ref{tbl-1}.
The lines were
measured with a K-band HEMT (high electron mobility transistor)
dual channel receiver, yielding spectra with angular resolutions of
$40''$ (NH$_3$) and $50''$ (HC$_3$N) in two orthogonally
oriented linear polarizations. Averaging the emission from both
channels, typical system temperatures are 100-150\,K for NH$_3$
and 80--100\,K for HC$_3$N on a main beam brightness temperature scale.

The measurements were carried out in frequency switching mode
using a frequency throw of $\sim$5\,MHz. The backend was an FFTS
(fast fourier transform spectrometer), operated with its minimum
bandwidth of 20\,MHz, providing simultaneously 16\,384 channels
for each polarization. The resulting channel widths are 15.4 and
20.1 \ms\ for NH$_3$ and HC$_3$N, respectively. We note,
however, that the true velocity resolution is about twice as large.

Observations started by measuring the continuum emission of
calibration sources (NGC\,7027, W3(OH), 3C\,286) and continued by
performing pointing measurements toward a source close to the spectroscopic
target. Spectral line measurements were interspersed with pointing
measurements at least once per hour. The calibration is estimated
to be accurate to $\pm$15\% and the pointing accuracy to be superior to
10\,arcsec.
The CLASS reduction package was used for standard data reduction.

{\it Nobeyama.}\, The NH$_3$ $(J, K) = (1, 1)$ line
at 23 GHz and the N$_2$H$^+$ $J$=1--0 line at 93 GHz
were observed with the Nobeyama Radio Observatory (NRO)
45-m telescope\footnote{The 45-m radio telescope is operated by
Nobeyama Radio Observatory, a brach of the National Astronomical Observatory
of Japan.} 
on April 8-10, 2009.  
We used a low-noise HEMT receiver, H22,
for the NH$_3$ observations and the two sideband-separating SIS
(Superconductor-Insulator-Superconductor)
receiver, T100 (Nakajima \etal\ 2008), for the N$_2$H$^+$ observations.
Both of them are dual polarization receivers. We observed two
polarizations simultaneously.  The 45-m radio telescope angular resolution
is about 73 and 17 arcsec at 23 and 93 GHz, respectively,
and the main beam efficiency is 0.84 at 23 GHz, and 0.53 at 93 GHz.
Autocorrelators were employed as a backend with bandwidth and 
channel separation
of 4 MHz and 4.375 kHz, respectively.
This corresponds to channel widths of
57 \ms\ at 23 GHz, and 14 \ms\ at 93 GHz.

The telescope pointing was checked by observing nearby SiO maser
sources every 1--2 hr. The pointing accuracy was $\leq 5''$.  
The line intensities were calibrated using the
chopper wheel method, and the observations were carried out 
in position switching mode.
The data reduction was performed partly with the AIPS-based software
package NewStar developed at NRO.

\section{Data analysis}
\label{sect-4}

\subsection{$V_{lsr}$ calculation}
\label{sect-4-1}

The radial velocities, $V_{lsr}$, are determined from the spectral line analysis.
Each individual exposure was first visually analyzed and corrupted data were excluded. 
Individual exposures were then coadded to increase the signal-to-noise
(S/N) ratio. Before coadding, a baseline was removed from each spectrum. To define the baseline,
spectral intervals without emission lines and/or noise spikes 
were selected and the mean signal $T_i$ along with
its rms uncertainty $\sigma_i$ were calculated for each interval. 
A set of pairs $\{ T_i, \sigma_i \}$    
was used to find a baseline (regression line) by minimizing  $\chi^2$.
This baseline was typically linear but was in some cases quadratic or cubic.
Since individual rms uncertainties $\sigma_i$ were of the same order of magnitude,
their mean value $\sigma$ was assigned to the whole spectrum. 
The resulting spectra were coadded with weights that are inversionally proportional to their
variances, $\sigma^2$.

The line parameters such as the total optical depth in the transition, 
$\tau$, the radial velocity, $V_{lsr}$,
the line broadening Doppler parameter, $b$, and the amplitude, $A$, 
were estimated by fitting
a one-component Gaussian model to the observed spectra. The model was defined by
\begin{equation}
T(v) = A\cdot \left[ 1 - e^{-t(v)} \right]\, ,
\label{eq1}
\end{equation}
where
\begin{equation}
t(v) = \tau\cdot \sum^k_{i=1}\, a_i\, \exp\left[ -
\frac{(v - v_i - V_{lsr})^2}{b^2} \right]\, ,
\label{eq2}
\end{equation}
where $a_i, v_i$ is, respectively, the relative intensity of the $i$th hyperfine component and its
velocity separation from the reference frequency. 
The sum in Eq.~(\ref{eq2}) runs over the $k$ 
hyperfine structure (hfs) components of the transition. 
The physical parameters $a_i$, and $v_i$ for the \amm\ (1,1), \hcccn\ (2--1), and \nnhp\ (1--0) transitions
are listed in Tables~\ref{tbl-2}, \ref{tbl-3}, and \ref{tbl-4}, respectively.  

The fitting parameters $\{A, \tau, b, V_{lsr} \}$ 
were determined by means of a $\chi^2$-minimization procedure.
For optically thin transitions, Eq.(\ref{eq1}) transforms into 
\begin{equation}
T(v) = A\cdot t(v)\, ,
\label{eq3}
\end{equation}
which prevents the independent estimation of $A$ and $\tau$. 
In this case, the model parameters were $\{A\cdot \tau, b, V_{lsr} \}$.

Since we are mostly interested in the model parameters $V_{lsr}$ and $b$, their values are listed in 
Tables~\ref{tbl-5} and \ref{tbl-6}. The $1\sigma$
errors of $V_{lsr}$ and $b$ were estimated from the diagonal elements of the covariance matrix
calculated for the minimum of $\chi^2$. 
The error in $V_{lsr}$ was also estimated independently by the
$\Delta \chi^2$ method (e.g., Press \etal\ 1992) to control both results.
When two estimates differed, the larger error was adopted.

\subsection{Uncertainties in the rest-frequencies}
\label{sect-4-2}

\begin{figure*}[t]
\vspace{0.0cm}
\hspace{-1.2cm}\psfig{figure=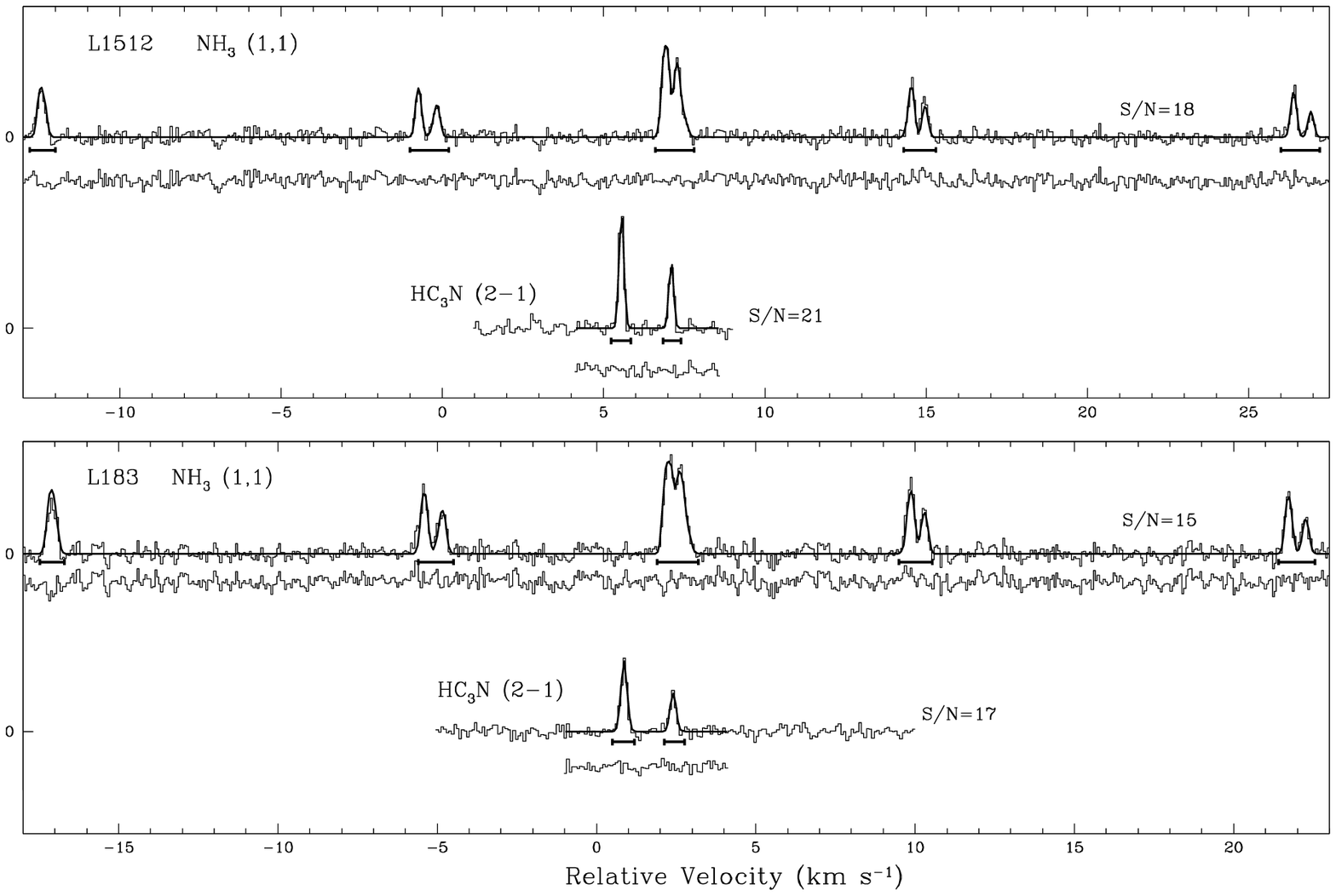,height=18cm,width=20cm}
\vspace{-7.0cm}
\caption[]{Spectra of NH$_3$ (1,1) and HC$_3$N ($2-1$) toward the cores L1512
and L183 obtained at the Medicina 32-m radio telescope.
The histogram shows the data, the solid curve shows the fit, and the residual
is plotted below each profile. 
The horizontal thick bars mark spectral windows used in the fitting procedure.
The data are the arithmetic means of all the observations.
The size of the resolution element (pixel) is 62 \ms\ for \amm, and
80 \ms\ for \hcccn. For each spectrum, the signal-to-noise ratio (S/N)
per pixel at the maximum intensity peak is depicted.
}
\label{fg2}
\end{figure*}

Molecular lines observed in cold dark clouds of $T_{\rm kin} \sim 10$K can
be extremely narrow with a thermal broadening of $v_{\rm th} \la 100$ \ms\, for \amm\,
or even lower for heavier molecules.
They provide a sensitive spectroscopic measurement of shallow 
velocity gradients in molecular clouds (Lapinov 2006).
At high spectral resolution ($FWHM \sim 30$ \ms), the uncertainty in the line position 
measurement from radio astronomical observations can be
as small as 1 \ms\, which is comparable to the precision 
available for laboratory rest-frame frequencies of the \amm\, $(J,K) = (1,1)$ transition
(Kukolich 1967; Hougen 1972). 
The physical parameters of the observed hfs components of \amm\, are listed in
Table~\ref{tbl-2}. Column 1 indicates the group numbers shown in the upper panel of Fig.~\ref{fg2}.
The errors in the line positions given in parentheses in Col.~7 are
smaller than 1 \ms\, with a mean of $\varepsilon_v = 0.58$ \ms.   
However, laboratory uncertainties for other molecules used in the present observations
are significantly larger.

The analysis of all available laboratory data about HC$_3$N carried out
by M\"uller \etal\ (2005) and independently by Lapinov (2008, private comm.)
shows good agreement,
the latter results being of slightly higher precision ($\varepsilon_v \simeq 2.8$ \ms).
The physical parameters of the hfs components of the $J = 2-1$ transition
are presented in Table~\ref{tbl-3}. The line identification numbers from Col.~1
are depicted in the middle panel of Fig.~\ref{fg2}.

The rest-frame frequencies for the third molecule N$_2$H$^+$ (Table~\ref{tbl-4})
were taken from 
the Cologne Database for Molecular Spectroscopy (CDMS) described in
M\"uller \etal\ (2005).
The group numbers from Table~\ref{tbl-4}, Col.~1 are also shown in Fig.~\ref{fg2},
lower panel.
The CDMS data have a factor of two smaller errors than
the hyperfine frequencies estimated from observations
of the molecular core L1512 in both N$_2$H$^+$ (93 GHz)
and C$_3$H$_2$ (85 GHz). 
It is again assumed 
that these molecules are co-spatially distributed (Caselli \etal\ 1995).
The error in $\varepsilon_v \simeq$13.5 \ms\, ($\simeq4$ kHz)
reported in Table~\ref{tbl-4} corresponds to the CDMS data.
The mean offset between the data from Caselli \etal\  and the CDMS is
$\Delta V = V_{\rm Caselli} - V_{\rm CDMS} = 12$ \ms. 

We do not correct the \nnhp\ frequencies for a 40 \ms\ offset as by
Pagani \etal\ (2009), who used the rest-frame frequencies of Caselli \etal. 
The offset of 40 \ms, or 28 \ms\ when adopting the CDMS rest frequencies (Molaro \etal\ 2009),
was determined by comparing the inversion
\amm\ (1,1) and rotational \nnhp\ ($1-0$) transitions observed in the target L183; 
this correction is justifable only when \dmm $\equiv 0$.
Reliable calibrations of the hyperfine transition frequencies of 
\nnhp\ can be performed only by a high precision laboratory measurement. 

We note in passing that the estimate of the mean radial velocity based on the simultaneous
fitting of $n$ hyperfine transitions should be more precise than that based on
a single line. However, the improvement is not as high as $1/\sqrt{n}$ because the
relative positions of the individual hfs transitions are correlated. If we consider
$n$ velocity differences $\{v_1 - v_0$, $v_2 - v_0$, $\ldots$, $v_n - v_0 \}$,
where $v_0$ is the reference velocity, then it is easy to show that 
the correlation coefficient $\kappa_{i,j}$ between two of them ($i\neq j$) is given by  
\begin{equation}
\kappa_{i,j} = \frac{1}{\sqrt{(1+s^2_i)(1+s^2_j)}}\, ,
\label{am7}
\end{equation}
where $s_i = \sigma_{v_i}/\sigma_{v_0}$ and $s_i = \sigma_{v_j}/\sigma_{v_0}$.
Taking into account that the laboratory errors in the hyperfine line positions,
$\sigma_{v_i}$,
are almost equal, we have $\kappa_{i,j} = \kappa \approx 1/2$.

The covariance matrix $Cov(v_i - v_0, v_j - v_0)$ 
contains $n$ diagonal terms $\sigma^2$ and $n(n-1)$ 
non-diagonal terms $\kappa\sigma^2$, where $\sigma^2$ represents the variance
in a single measurement.
The error in the mean radial velocity caused by the laboratory
uncertainties (referred to as $\varepsilon_{\rm sys}$ hereafter) 
can be calculated as described by, e.g., Stuart \& Ord (1994)
\begin{equation}
\varepsilon_{\rm sys} = \left[ \sum^n_{i=1}\sum^n_{j=1} \omega_i\omega_j 
Cov(v_i - v_0, v_j - v_0) \right]^{1/2}\, .
\label{am8}
\end{equation}
In cases of equal accuracy, the weight $\omega_i = 1/n$ for each $i$.
Then $\varepsilon_{\rm sys}$ is approximately equal to
\begin{equation}
\varepsilon_{\rm sys} = \frac{\sigma}{n}\sqrt{n + n(n-1)\kappa} \approx \sigma\sqrt{\kappa}\ .
\label{am9}
\end{equation}
Thus, the gain factor, $\sqrt{\kappa}$, is only about 0.7 for the molecules in question.

\section{Results}
\label{sect-5}

\begin{figure*}[t]
\vspace{0.0cm}
\hspace{0.0cm}\psfig{figure=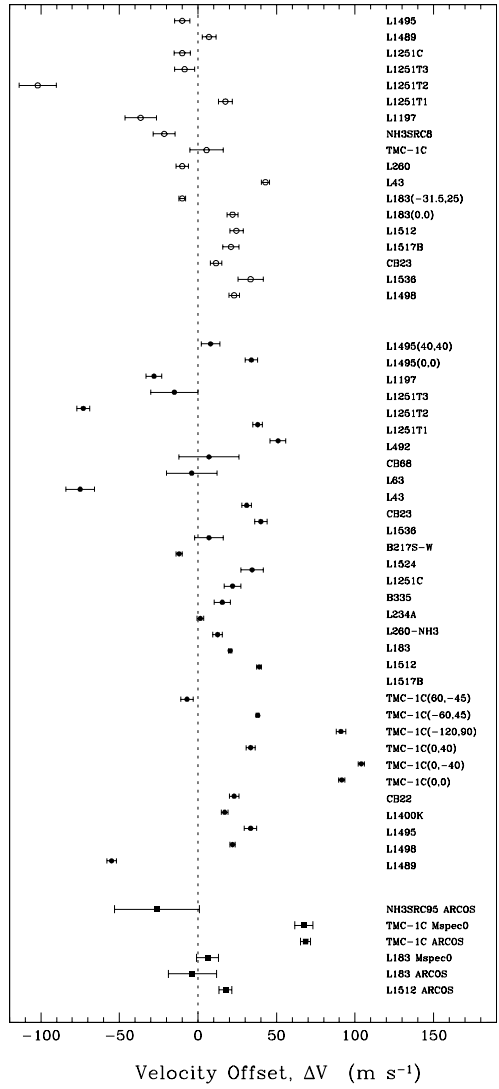,height=15cm,width=15cm}
\vspace{-0.1cm}
\caption[]{
Doppler velocity differences, $\Delta V$, between the \hcccn\ $J=2-1$ and \amm\ $(J,K) = (1,1)$
transitions for the data obtained at the 32-m (filled squares) and the 100-m (filled circles)
telescopes, and between
the \nnhp\ $J=1-0$ and \amm\ $(J,K) = (1,1)$ transitions
for the 45-m telescope data (open circles). The 1$\sigma$ statistical errors are indicated.
Given in parentheses are the coordinate offsets in arcsec.
}
\label{fg10}
\end{figure*}

We first consider the entire sample of sources listed in Table~\ref{tbl-1}.   
By applying the same computational procedure to all available spectra from the three
radio telescopes, we find that not all of the molecular profiles can be described adequately
with a single-component Gaussian model. 
In total, we measured $n = 55$ molecular pairs. 
The corresponding velocity offsets $\Delta V$ 
are shown in Fig.~\ref{fg10} by three types of symbols: 
filled squares and circles (\amm/\hcccn\ pairs), and open circles 
(\amm/\nnhp\ pairs) are, respectively, data points from the 32-m, 100-m, and 45-m telescopes. 
Some molecular cores
(L183, L1495, TMC-1C) were  partially mapped.  
The offset coordinates in arcsec with respect to the source positions from Table~\ref{tbl-1} 
are shown in parentheses. Two objects (L183 and TMC-1C) were observed
at the 32-m telescope with different spectrometers ARCOS and MSpec0, as indicated in the figure. 
Five sources, TMC-1/HC3N (Medicina), TMC-2, L1521F, and L1544 (Effelsberg), and
TMC-2 (Nobeyama) from Table~\ref{tbl-1}
exhibit asymmetric profiles that are not consistent with 
a simple Gaussian model (we do not show them in Fig.~\ref{fg10}). 
A considerable fraction of the molecular pairs from the total sample
exhibit non-thermal motion, i.e., the Doppler broadening parameter
$b$(\amm) $\leq b$(\hcccn, \nnhp).  

Several sources were observed at different radio telescopes.
Since these data points have different systematic errors, 
we treat them as `independent' measurements
in the following statistical estimates. 

The weighted mean (weights inversionally proportional to the variances) 
of the ensemble of $n=55$ $\Delta V$ values is 
$\langle \Delta V \rangle _{\scriptscriptstyle \rm W} 
= 27.4\pm4.4$ \ms, the scale (standard deviation) is 32.6 \ms,
and the median is 17 \ms.
We also used a robust redescending $M$-estimate for the mean
and the normalized median absolute deviation ($1.483\cdot$MAD) for the scale.
These statistics work well for inhomogeneous data sets
with outliers and deviations from normality
(corresponding formulae are given in Appendix).
The $M$-estimate infers that
$\langle \Delta V \rangle _{\scriptscriptstyle \rm M} 
= 14.1\pm4.0$ \ms\ (scale 29.6 \ms). 
A poor concordance between three mean estimates 
($\langle \Delta V \rangle _{\scriptscriptstyle \rm W}$,
$\langle \Delta V \rangle _{\scriptscriptstyle \rm M}$, and the median)
is caused by large systematic shifts and, as a result,
by `heavy tails' of the probability distribution function. 
The scatter in the points reflects effects related to the gas kinematics
and the chemical segregation of one molecule with respect to
the other.

An additional decrease in the noise component 
$Var(\Delta V_n)$ in Eq.~(\ref{am6}) 
is possible if we select from the sample of the observed
targets the systems with `simple' geometry and internal kinematics. 
The ideal target would be
a homogeneous spherical cloud where different molecules are co-spatially distributed 
(no chemical segregation) and where turbulence is suppressed (thermally
dominated motion).  In this case, any deviations from the expected zero mean value
of the radial velocity difference between rotational and inversion molecular
transitions in Eq.~(\ref{am4}) could be ascribed to the non-zero $\Delta \mu$ value.

In practice, molecular cores are
not ideal spheres and when observed at higher angular resolutions
they frequently exhibit complex substructures.
The line profiles may be asymmetric because of non-thermal bulk motions. 
Taking this into account, the following criteria were formulated:
\begin{enumerate}
\item[1.]
The line profiles are symmetric described well by a single-component Gaussian model
(i.e., the minimum value of $\chi^2_\nu \sim 1$).
This selection increases the accuracy of the line center measurement.
Multiple line components may shift the line barycenter and affect the velocity
difference between molecular transitions because, e.g., the ratio \amm/\hcccn\ can vary
from one component to another.
\item[2.]
The line widths do not greatly exceed the Doppler width because of the thermal motion of material,
i.e., the non-thermal component (infall, outflow, tidal flow, turbulence) does not dominate 
the line broadening. This ensures that selected molecular lines correspond to the same
kinetic temperature and arise cospatially. 
For the molecules in question we require that the ratio of the Doppler $b$-parameters,
$\beta = b$(\amm)$/b$(\hcccn) or $\beta = b$(\amm)$/b$(\nnhp), be $\beta \geq 1$.  
\item[3.]
The spectral lines are sufficiently narrow ($b \sim 0.1-0.2$ \kms) 
for hyperfine structure components to be resolved. 
This allows us to validate 
the measured radial velocity by means of different hfs lines of the same molecular transition.
\item[4.]
The spectral lines are not heavily saturated and their profiles are not
affected by optical depth effects.
The total optical depth of the \amm\ hf transitions is $\tau \leq 10$. 
\end{enumerate}

We now consider molecular lines that fulfill these additional selection criteria. 
Among the sources listed in Table~\ref{tbl-1}, two and twelve molecular 
cores can be selected from, respectively, the Medicina and Effelsberg 
observations (Table~\ref{tbl-5}), and nine clouds from the Nobeyama
data set (Table~\ref{tbl-6}). The sample size is now $n=23$,
i.e., two times smaller than the total data set. 

\begin{figure*}[t]
\vspace{0.0cm}
\hspace{-0.2cm}\psfig{figure=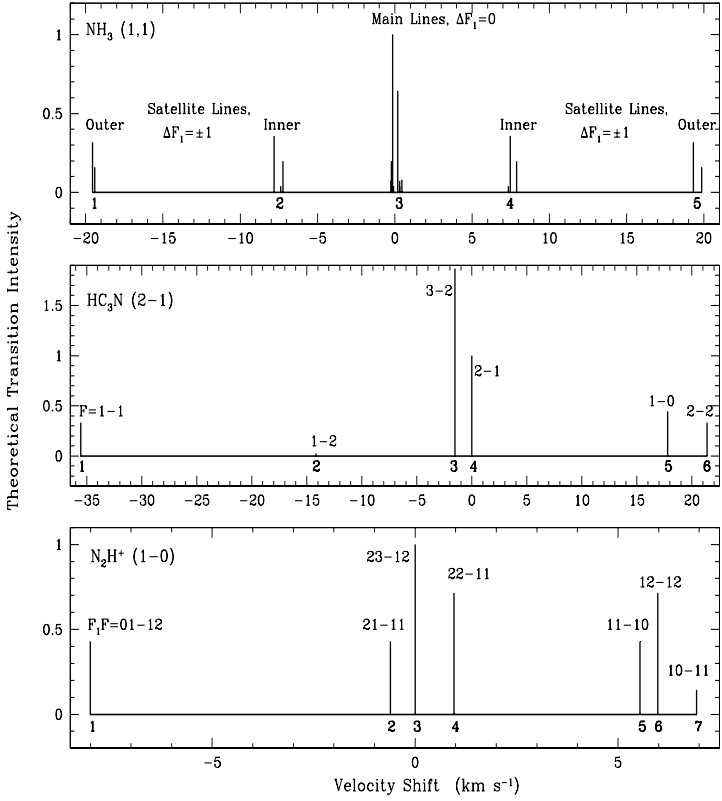,height=12cm,width=10cm}
\vspace{-1.0cm}
\caption[]{Sketch of the hyperfine transitions for the NH$_3$ (1,1), 
HC$_3$N $(2-1)$, and N$_2$H$^+$ $(1-0)$ states.
Line identification numbers are given in each panel (see Tables \ref{tbl-2}~--~\ref{tbl-4}). }
\label{fg1}
\end{figure*}

Table~\ref{tbl-5} lists the values obtained from the analysis of the \amm\
(1,1) and \hcccn\ (2--1) transitions, whereas in Table~\ref{tbl-6} the \amm\ (1,1) 
and \nnhp\ (1--0) measurements are presented. 
Where individual hfs transitions can be analyzed separately, the results of these
analyse are also given. 
The data obtained for \amm\ are divided formally into 4 groups, which are marked 
in Fig.~\ref{fg1} and Table~\ref{tbl-2} by the following numbers: `outer'~--  1 and 5,
`inner'~--2 and 4, `central'~-- 3, and `total' combines all hfs lines (Cols.~2-5 of Table~\ref{tbl-5},
respectively).
For \hcccn, the measurements are presented in 3 groups, which are marked in 
Fig.~\ref{fg1} and Table~\ref{tbl-3} as: `low'~-- 1, 5, 
and 6, `high'~-- 3 and 4, and `total'~--1, 3, 4,
5, and 6 (Cols.~6-8 of Table~\ref{tbl-5}, respectively). 
Line number 2 ($F=1-2$) is very weak and was never detected in our
observations.
For \nnhp, we divided the hfs lines into 3 groups on the basis 
of their relative theoretical intensities. 
These groups are indicated in 
Fig.~\ref{fg1} and Table~\ref{tbl-4} as: `low'~-- 1, 2, 5, and 7, `high'~-- 3, 4, and 6,
and `total' combines all hfs lines
(Cols.~6-8 of Table~\ref{tbl-6}, respectively). 

\begin{figure*}[t]
\vspace{0.0cm}
\hspace{-1.2cm}\psfig{figure=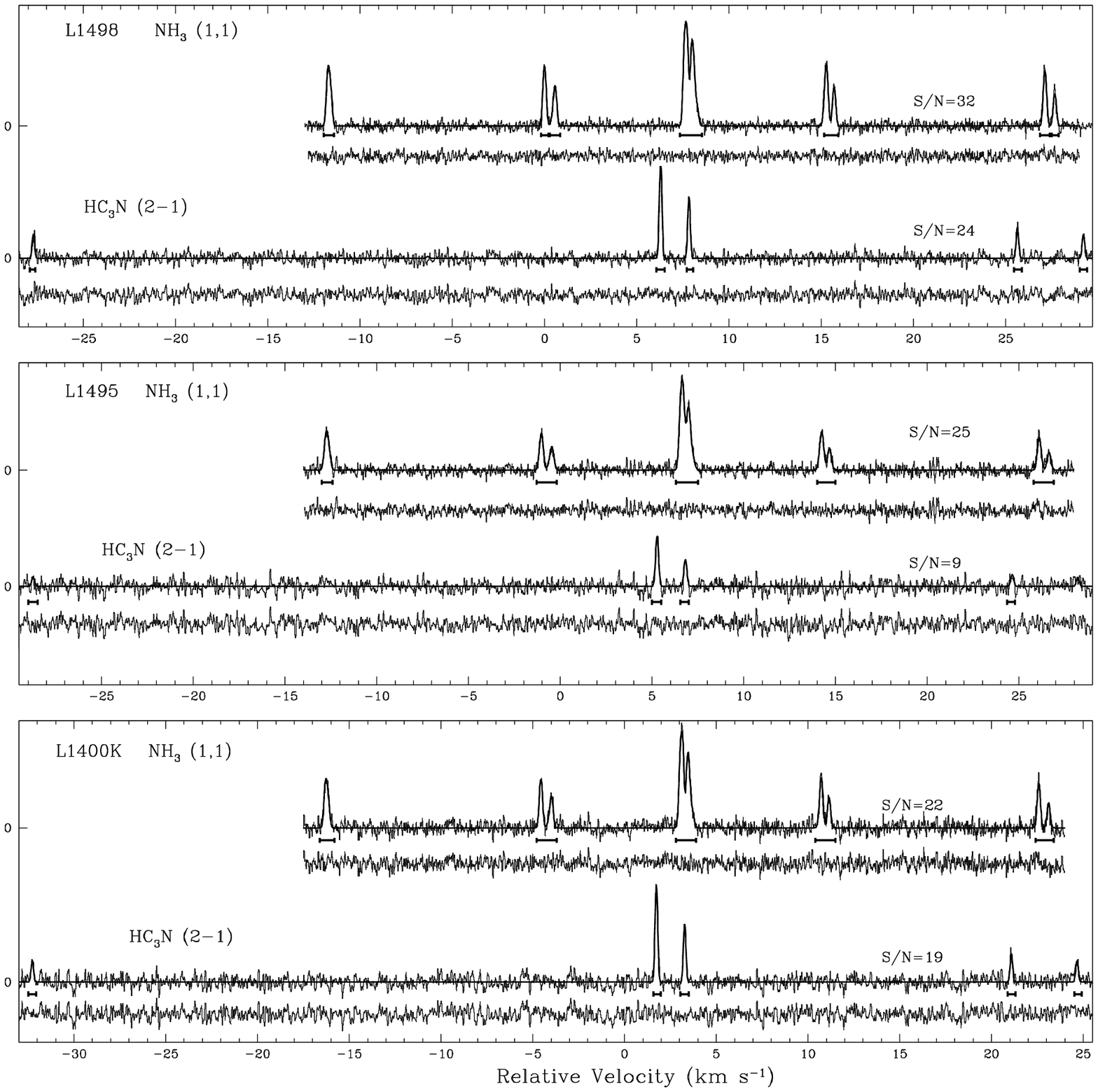,height=18cm,width=20cm}
\vspace{-1.5cm}
\caption[]{Spectra of NH$_3$ (1,1) and HC$_3$N ($2-1$) toward the cores L1498, L1495,
and L1400K obtained at the Effelsberg 100-m radio telescope.
The histogram shows the data, the solid curve shows the fit, and the residual
is plotted below each profile. 
The horizontal thick bars mark spectral windows used in the fitting procedure.
The data are the arithmetic means of all the observations.
The size of the resolution element (pixel) is 15 \ms\ for \amm, and
20 \ms\ for \hcccn. For each spectrum, the signal-to-noise ratio (S/N)
per pixel at the maximum intensity peak is depicted.
}
\label{fg3}
\end{figure*}

\begin{figure*}[t]
\vspace{0.0cm}
\hspace{-1.2cm}\psfig{figure=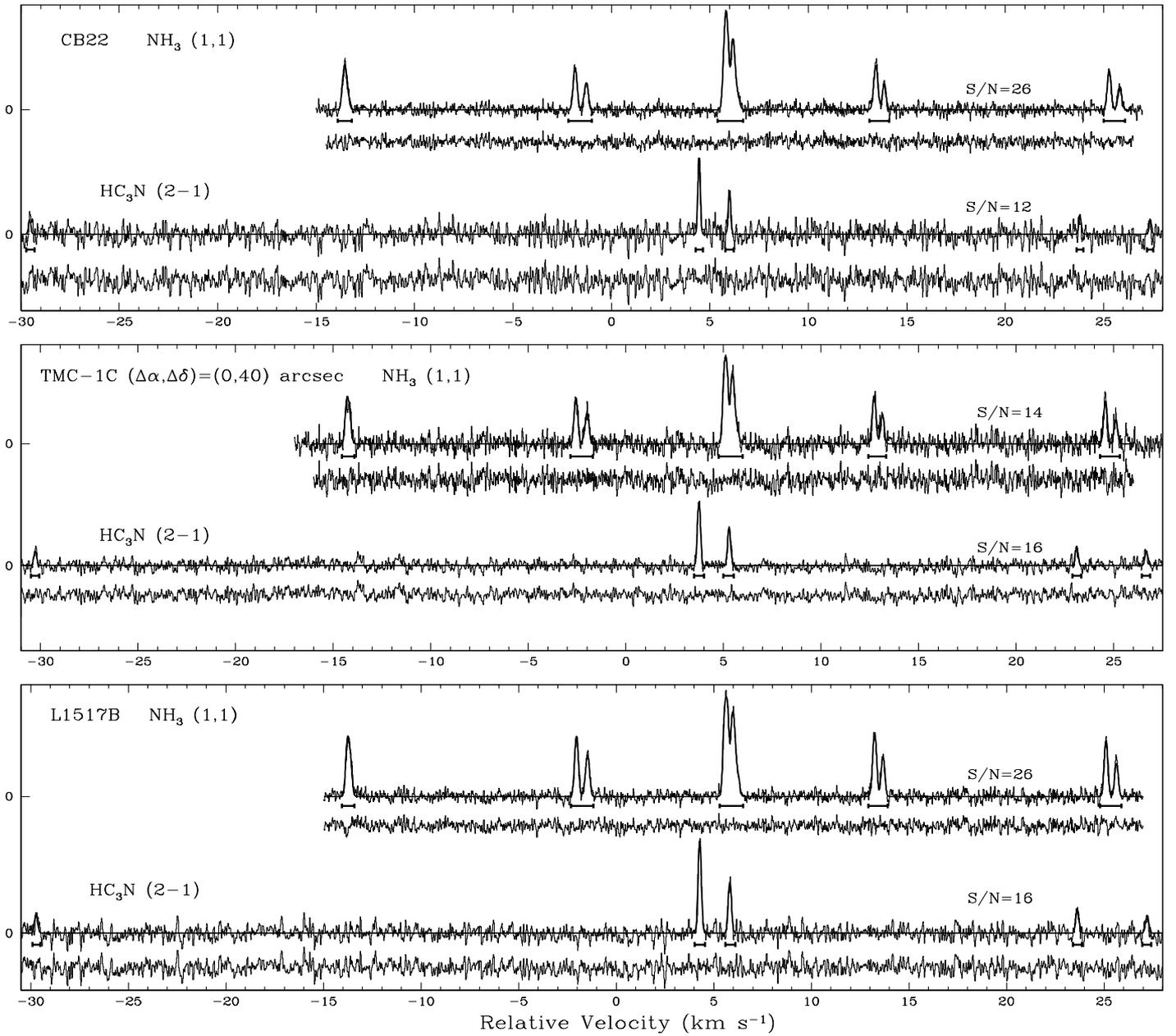,height=18cm,width=20cm}
\vspace{-1.5cm}
\caption[]{Same as Fig.~\ref{fg3} but for the cores CB22, 
TMC-1C (offset $\Delta\alpha,\Delta\delta = 0,40$
arcsec),  and L1517B. }
\label{fg4}
\end{figure*}

\begin{figure*}[t]
\vspace{0.0cm}
\hspace{-1.2cm}\psfig{figure=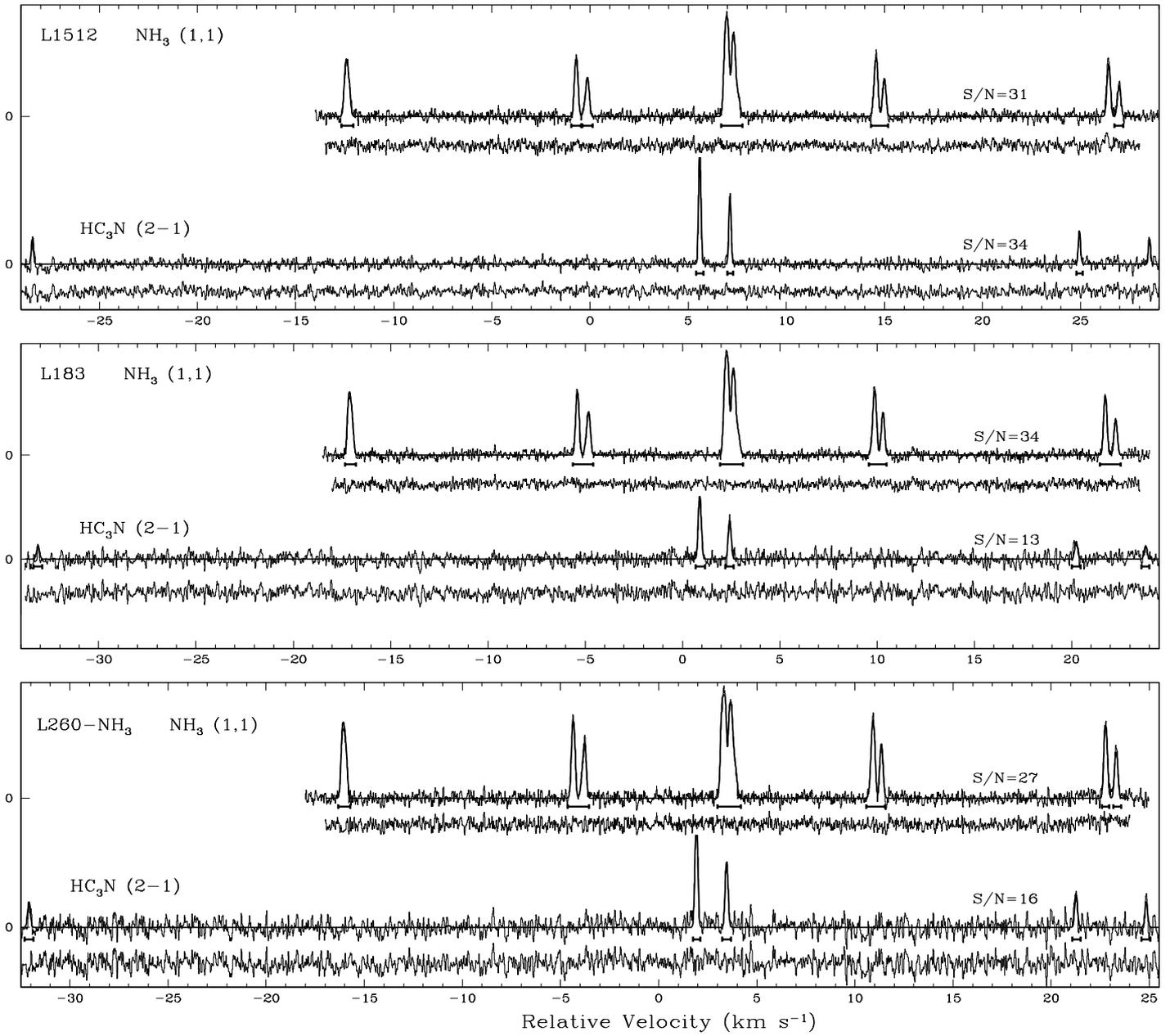,height=18cm,width=20cm}
\vspace{-1.5cm}
\caption[]{Same as Fig.~\ref{fg3} but for the cores L1512, L183, and L260-\amm. }
\label{fg5}
\end{figure*}

\begin{figure*}[t]
\vspace{0.0cm}
\hspace{-1.2cm}\psfig{figure=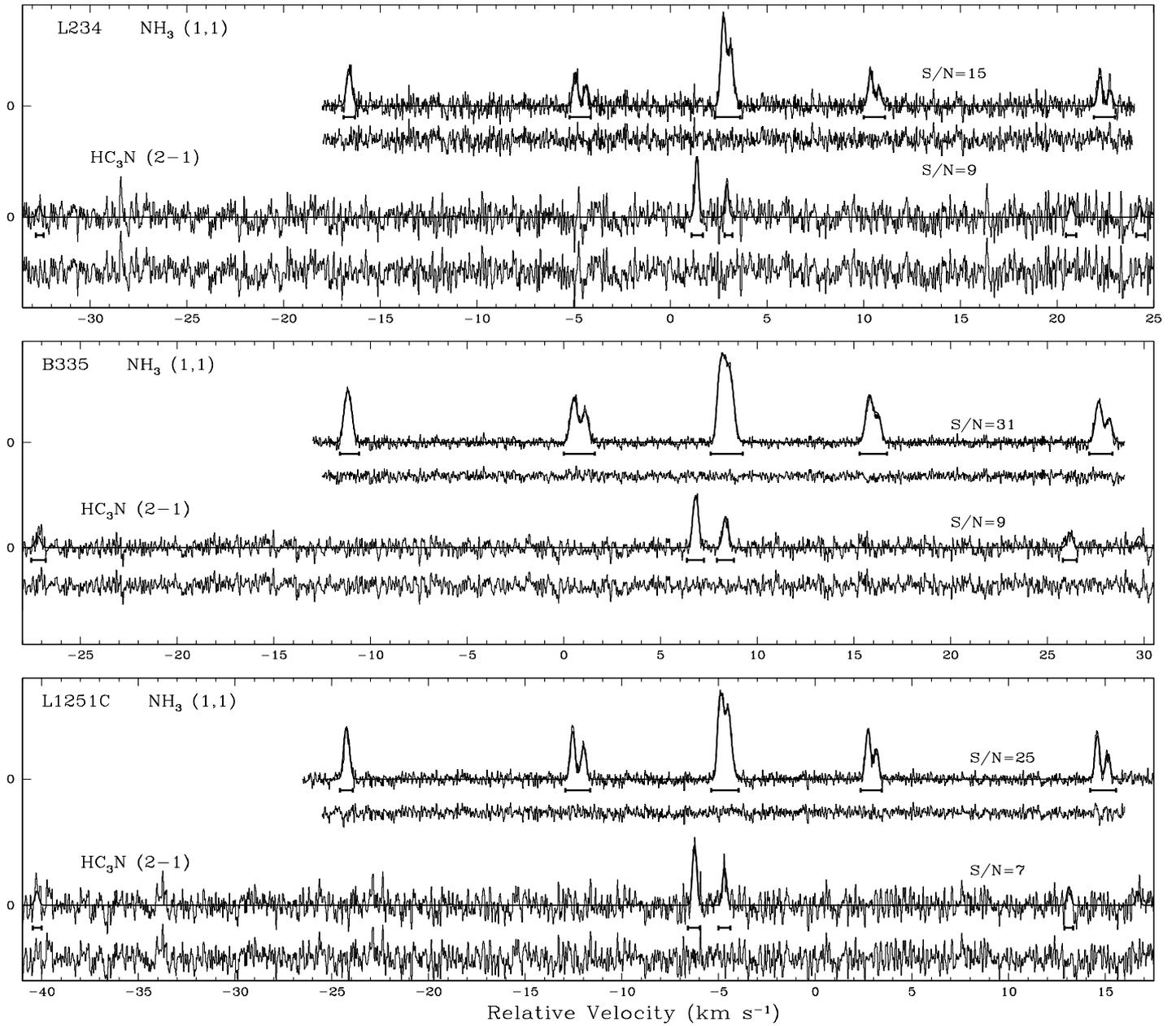,height=18cm,width=20cm}
\vspace{-1.5cm}
\caption[]{Same as Fig.~\ref{fg3} but for the cores L234A, B335, and L1251C. }
\label{fg6}
\end{figure*}

\begin{figure*}[t]
\vspace{0.0cm}
\hspace{-1.2cm}\psfig{figure=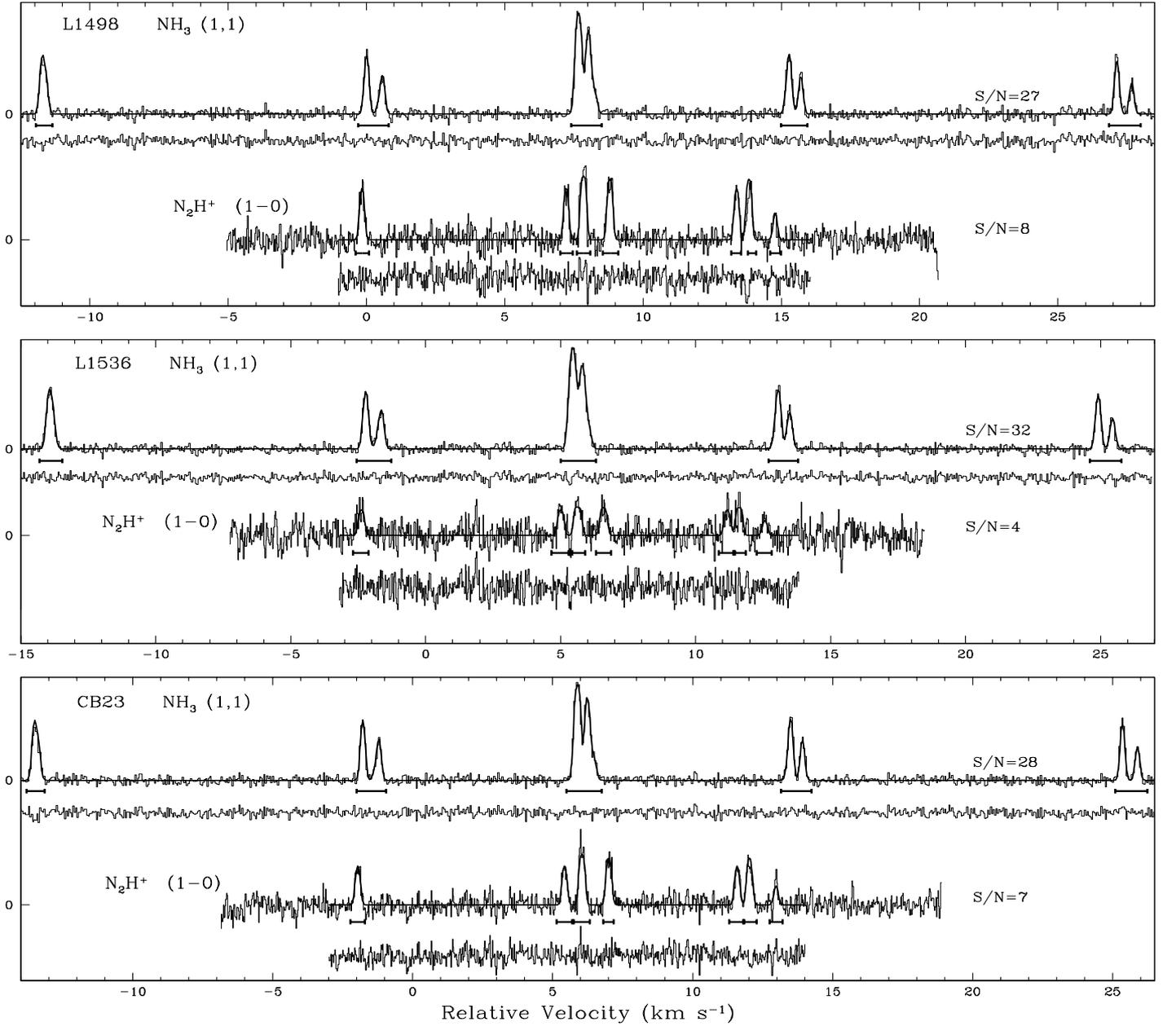,height=18cm,width=20cm}
\vspace{-1.5cm}
\caption[]{Spectra of NH$_3$ (1,1) and N$_2$H$^+$ ($1-0$) toward the cores L1498, L1536,
and CB23 obtained at the Nobeyama 45-m radio telescope.
The histogram shows the data, the solid curve shows the fit, and the residual
is plotted below each profile. 
The horizontal thick bars mark spectral windows used in the fitting procedure.
The data are the arithmetic means of all the observations.
The size of the resolution element (pixel) is 49 \ms\ for \amm, and
25 \ms\ for \nnhp. For each spectrum, the signal-to-noise ratio (S/N)
per pixel at the maximum intensity peak is depicted.
}
\label{fg7}
\end{figure*}

\begin{figure*}[t]
\vspace{0.0cm}
\hspace{-1.2cm}\psfig{figure=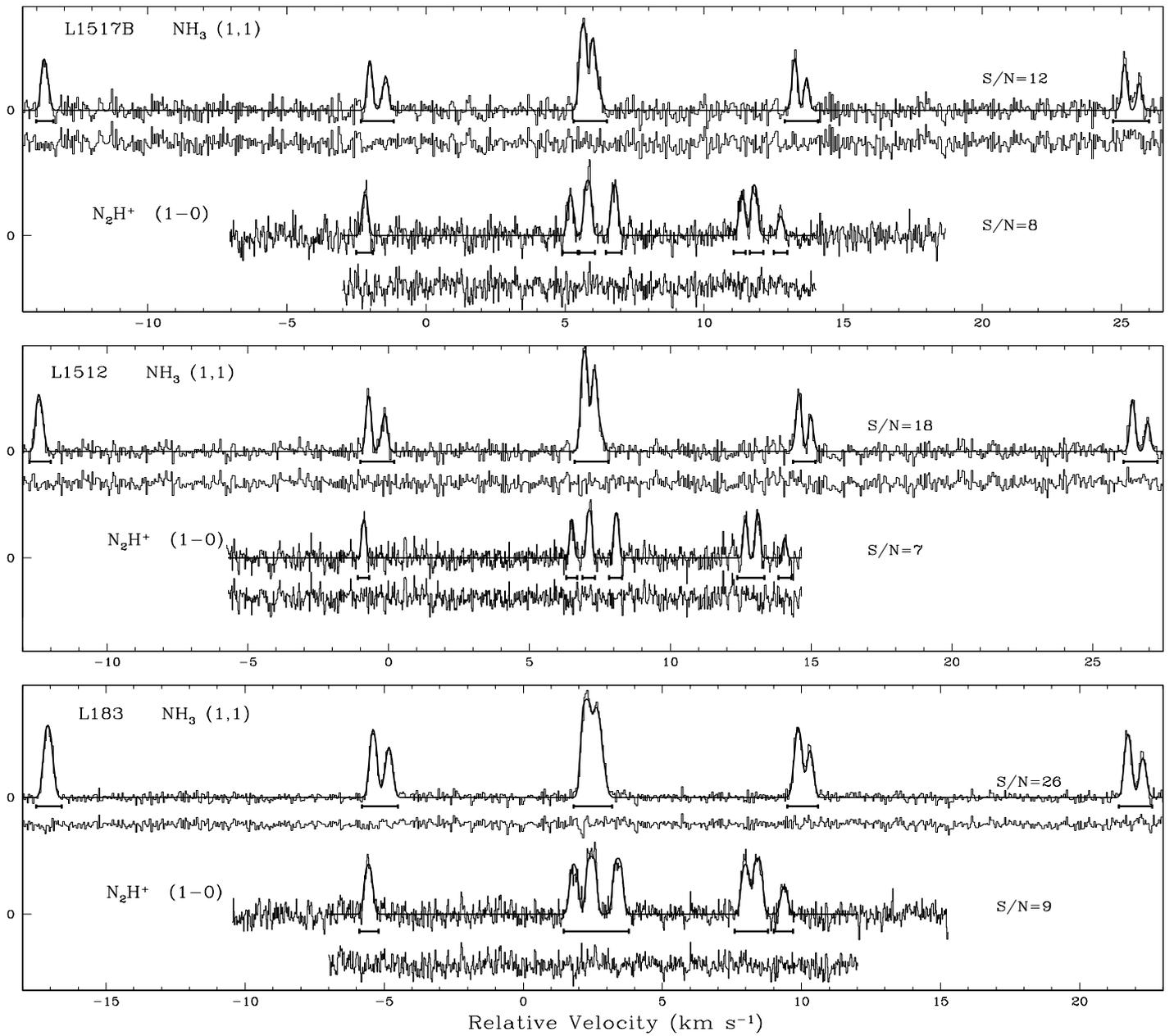,height=18cm,width=20cm}
\vspace{-1.5cm}
\caption[]{Same as Fig.~\ref{fg7} but for the cores L1517B, L1512, and L183. }
\label{fg8}
\end{figure*}

\begin{figure*}[t]
\vspace{0.0cm}
\hspace{-1.2cm}\psfig{figure=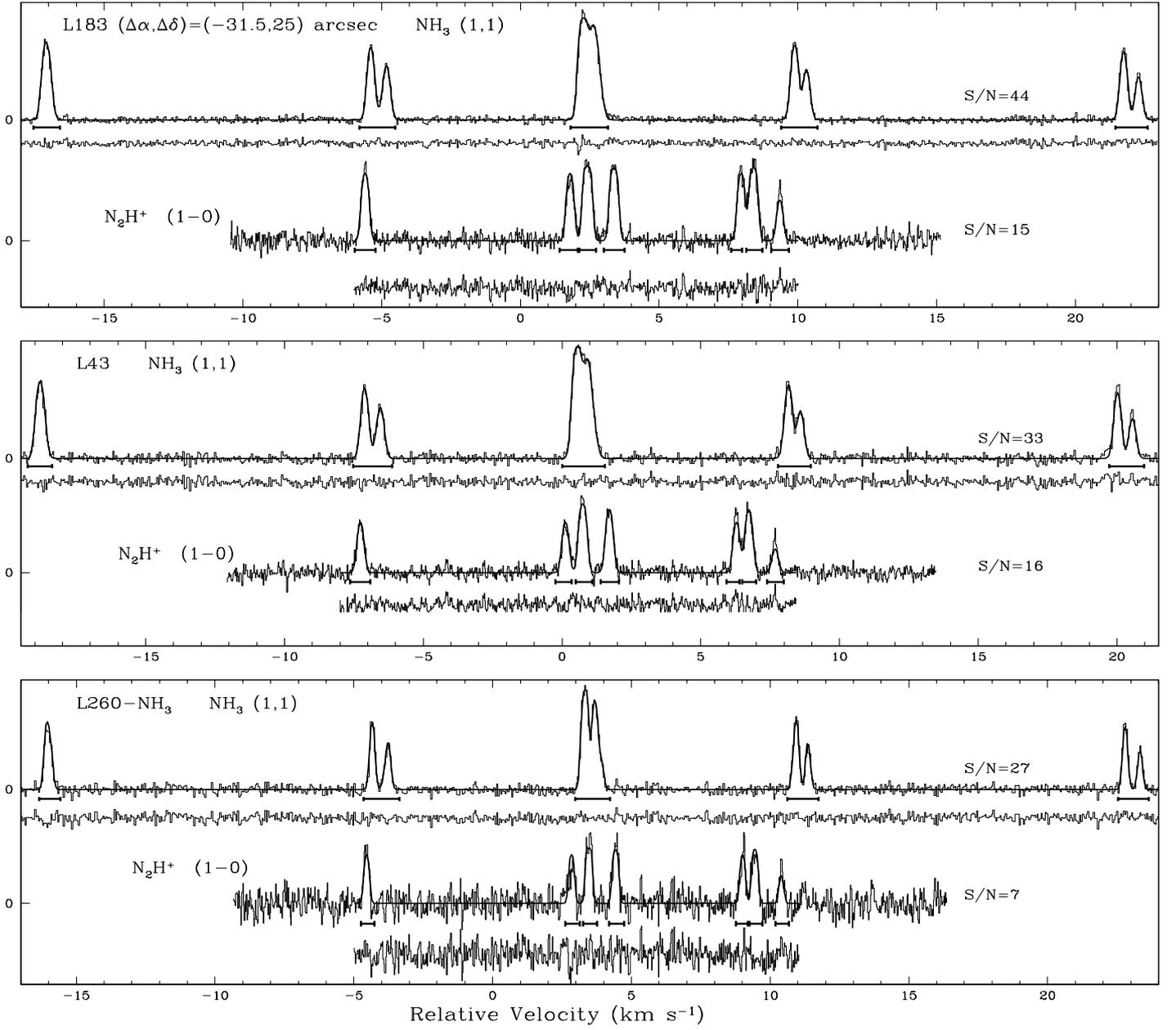,height=18cm,width=20cm}
\vspace{-1.5cm}
\caption[]{Same as Fig.~\ref{fg7} but for the cores L183 (offset $\Delta\alpha,\Delta\delta =
-31.5,25$ arcsec), L43, and L260-\amm. }
\label{fg9}
\end{figure*}

The analyzed molecular line profiles 
are shown in Figs.~\ref{fg2}-\ref{fg9}. The smooth curves are synthetic spectra calculated in
the simultaneous fit of all hyperfine components to the observed profiles. Bold horizontal lines mark 
the spectral ranges included in the $\chi^2$-minimization procedure. 
The quality of individual fittings can be characterized by the
normalized $\chi^2_{\nu}$ values given in Tables~\ref{tbl-5} 
and \ref{tbl-6} for each group of measurements. 
The residuals `observed data -- model' are depicted beneath 
each spectrum in Figs.~\ref{fg2}-\ref{fg9}.
The signal-to-noise ratio (S/N)  
per resolution element shown in these figures
is calculated at the maximum intensity peak.

\begin{figure*}[t]
\vspace{0.0cm}
\hspace{0.0cm}\psfig{figure=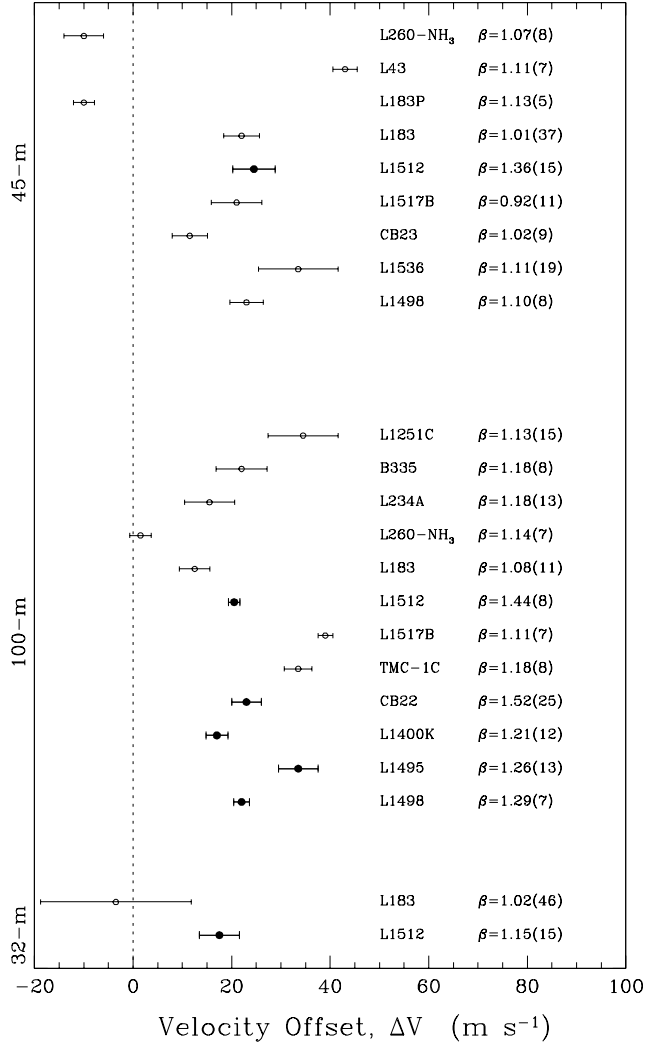,height=15cm,width=15cm}
\vspace{-0.8cm}
\caption[]{
Doppler velocity differences, $\Delta V$, between the \hcccn\ $J=2-1$ and \amm\ $(J,K) = (1,1)$
transitions for the data obtained at the 32-m and 100-m telescopes, and between
the \nnhp\ $J=1-0$ and \amm\ $(J,K) = (1,1)$ transitions
for the 45-m telescope data (1$\sigma$ statistical errors are indicated).
The parameter
$\beta$ is the ratio of the Doppler $b$-parameters: $\beta = b$(\amm)$/b$(\hcccn), or
$\beta = b$(\amm)$/b$(\nnhp). Given in parentheses are $1\sigma$ errors.  
The filled circles mark sources with thermally dominated motions.
}
\label{fg11}
\end{figure*}

The velocity offsets, $\Delta V$, and their statistical errors are listed in Col.~9 
of Tables~\ref{tbl-5} and \ref{tbl-6}. When calculating
$\Delta V$, we used the mean $V_{lsr}$ radial velocities based on the 
simultaneous fitting of all hyperfine transitions. 
The $\Delta V$ values estimated in 23 measurements
are marked in Fig.~\ref{fg11} by the circles with $1\sigma$ error bars.
The filled circles indicate sources with thermally dominated motions.
We also show the ratios of the Doppler $b$-parameters: $\beta = b$(\amm)$/b$(\hcccn) or
$\beta = b$(\amm)$/b$(\nnhp), in parentheses provide their $1\sigma$ errors.  
For the 32-m and 100-m telescopes, 
$\Delta V = V_{lsr}$(HC$_3$N) $- V_{lsr}$(NH$_3$), 
whereas for the 45-m telescope
$\Delta V = V_{lsr}$(N$_2$H$^+$) $- V_{lsr}$(NH$_3$). 

Among the selected clouds, we discover that only one (L260-NH$_3$) violates the selection rule No.~4.
The total optical depth in this case is $\tau = 15.6$. We used this cloud since the spectrum of \amm\
was of high S/N and other selection criteria were fulfilled. We note, however, that
this cloud has the minimum $\Delta V$ value consistent with a zero offset 
(see Table~\ref{tbl-5} and Fig.~\ref{fg11}). 

Both data sets exhibit an excess of positive velocity offsets.
The most accurate results are obtained for the L1498 and L1512 molecular 
cores observed with the 100-m Effelsberg
telescope, $\Delta V = 21.0\pm1.5$ \ms. 
In this case, the systematic error due to the rest frequency uncertainties does not exceed
3 \ms. The maximum spread between the individual hfs 
$V_{lsr}$ velocities for \amm\ is 5.5 \ms\ (L1498)
and 4.5 \ms\ (L1512), but it is only 1.0 \ms\ for the
\hcccn\  hyperfine transitions
in both sources.
The molecular lines are narrow with 
$b$(\amm) $= 118\pm1$ \ms,  $b$(\hcccn) $= 91\pm5$ \ms\ (L1498), and 
$b$(\amm) $= 113\pm1$ \ms,  $b$(\hcccn) $= 79\pm4$ \ms\ (L1512), 
which is in line with the assumption that in these
two clouds both molecules trace the same volume elements. 

The sources with thermally dominated motions ($n = 7$) marked 
by the filled circles in Fig.~\ref{fg11}
give the weighted mean of
$\langle \Delta V \rangle _{\scriptscriptstyle \rm W} 
= 21.1\pm1.3$ \ms, the scale 3.4 \ms, and the median 22 \ms.
The corresponding $M$-estimate is 
$\langle \Delta V \rangle _{\scriptscriptstyle \rm M} 
= 21.2\pm1.8$ \ms\ (scale 4.8 \ms).

For the ensemble of $n=23$ $\Delta V$ values, 
we found the weighted mean of
$\langle \Delta V \rangle _{\scriptscriptstyle \rm W} 
= 20.7\pm3.0$ \ms, the scale 14.4 \ms, and the median 22 \ms.
The robust $M$-estimate is 
$\langle \Delta V \rangle _{\scriptscriptstyle \rm M} 
= 21.5\pm2.8$ \ms\ (scale 13.4 \ms).
Thus, in the reduced sample, we have good agreement between all three
estimates of the mean.
The scatter in the points (the Doppler noise) is
lower by a factor of two than that of the $n=55$ data set. 

The individual data from the 100-m and 45-m telescopes provide
the following estimates.
{\it Effelsberg:}
$\langle \Delta V \rangle _{\scriptscriptstyle \rm W}
= 23.0\pm3.1$ \ms\ ($n=12$, scale 10.7 \ms, median 22 \ms), 
$\langle \Delta V \rangle _{\scriptscriptstyle \rm M} 
= 23.2\pm3.8$ \ms\ (scale 13.3 \ms).
{\it Nobeyama:}
$\langle \Delta V \rangle _{\scriptscriptstyle \rm W}
= 14.0\pm7.2$ \ms\ ($n=9$, scale 21.6 \ms, median 22 \ms),
$\langle \Delta V \rangle _{\scriptscriptstyle \rm M} 
= 22.9\pm4.2$ \ms\ (scale 12.7 \ms).

Although the robust $M$-estimates of the mean for both the Effelsberg 
and Nobeyama observations are consistent,
the latter has a larger systematic error due to its lower accuracy at the rest
frequencies of the \nnhp\ (1--0) transition. Taking into account that the rest
frequencies of the \hcccn\ (2--1) are known with a sufficiently high accuracy
(the uncertainties in the laboratory and observational frequencies are comparable),
we take the Effelsberg robust mean as a final value for the velocity offset between the
rotational and inversion transitions. Being interpreted in terms of the electron-to-proton
mass ratio variation, this provides the value
\dmm = $22\pm4_{\rm stat}\pm3_{\rm sys}$ ppb\ (1 ppb = $10^{-9}$).

\subsection{Data reproducibility}
\label{sect-5-1}

Taking into account numerous perturbation effects discussed above
and variations in specific parameters such as spectral resolution and
signal-to-noise ratio, we question the consistency
of the measured $\Delta V$ values
obtained for independent telescope systems. This consistency was first 
tested at the 32-m Medicina telescope, where
we observed two cores, L183 and TMC-1C, in lines of \amm\ (1,1) 
and \hcccn\ (2-1) with the ARCOS and MSpec0 digital spectrometers.
For L183, we found
$\Delta V = 6\pm7$ \ms\  (MSpec0) and $-4\pm15$ \ms\ (ARCOS),
whereas for TMC-1C, the corresponding quantities were measured to be
$67.5\pm5.8$ \ms\  and $68.5\pm3.2$ \ms.
Both results are in good agreement to within the $1\sigma$ uncertainty interval.

We also tested the reliability of the velocity offsets by obtaining observations
of the same cores at different telescopes.
For instance, a quiescent low-mass molecular core L1512 in the Taurus Cloud 
was observed in the \amm\ (1,1) and \hcccn\ (2--1) lines at
the 32-m and 100-m telescopes. The corresponding velocity offsets are
$\Delta V = 17.5\pm4.1$ \ms\ and $20.5\pm1.2$ \ms. Other examples can be
found in Fig.~\ref{fg10}, where velocity offsets $\Delta V$ of
molecular pairs consistent
with a singe-component Gaussian model are depicted.

Some of our targets were partially mapped as indicated in Fig.~\ref{fg10}.  
The six points of TMC-1C are scattered between
$\Delta V = -7\pm4$ \ms\ and $104\pm2$ \ms, which is indicative of large systematic
shifts caused by bulk motions. Indeed, the line widths of \amm\ and
\hcccn\ demonstrate the dominant influence of the non-thermal component.
The measured ratio $\beta = b$(\amm)/$b$(\hcccn) at these points is equal to
1.0, 0.8, 1.2, 1.0, 1.2, and 1.0 at, respectively,
$(\Delta \alpha, \Delta \delta) = (0,0), (0,-40), (0,40),
(-120,90), (-60,45)$, and $(60,-45)$ arcsec.
Similar velocity offsets are observed  for L1495 and L183.
In general, the scatter for the whole sample covers the range
$-100 \la \Delta V \la 100$ \ms.
However, the scatter for the subsample selected in accord
with the additional selection criteria from Sect.~\ref{sect-5} decreases considerably.
The corresponding points shown in Fig.~\ref{fg11} are distributed across the
interval from $-10$ to 43 \ms.
Thus, as expected, molecular cores with thermally dominated gas motions are
the most suitable targets for precise measurements of \dmm.

It is essential that the observed spectral-line displacements 
are based on differential measurements of radial velocities
recorded at the {\it same} telescope over short periods of time.
It is also important that different receivers used in these observations
enable the independent Doppler tracking of the observed molecular lines.
Thus, there is no need for an accurate definition of the zero point, and only
internal precisions of the measured
line positions and the uncertainties in the rest-frame frequencies
restrict the accuracy of the measured velocity offset.

\subsection{Comparison with previously obtained results}
\label{sect-5-2}

In our preliminary analysis (LMK), we used
high quality radio spectra of molecular cores
in lines of NH$_3$ $(J,K) = (1,1)$ (23 GHz), 
CCS $J_N = 2_1-1_0$ (22 GHz), HC$_3$N $J = 5-4$ (45 GHz),
and N$_2$H$^+$ $J = 1-0$ (93 GHz). 
The observations were carried out with the 100-m Green Bank Telescope (GBT) 
by Rosolowsky \etal\ (2008) for the Perseus molecular cloud, 
by Rathborne \etal\ (2008) for the Pipe Nebula,
and with the Nobeyama 45-m telescope by Sakai \etal\ (2008) for infrared
dark clouds (IRDCs). 

The most accurate estimates were obtained from carefully
selected subsamples of the NH$_3$/CCS pairs observed with high 
spectral resolution (FWHM = 25 \ms) in the
Perseus molecular cloud ($n = 21$) and the Pipe Nebula ($n = 8$):
$\Delta V = 52\pm7_{\rm stat}\pm14_{\rm sys}$ \ms\ and
$69\pm11_{\rm stat}\pm14_{\rm sys}$ \ms, respectively.
The analysis of the $n = 36$ NH$_3$/N$_2$H$^+$ pairs and
$n = 27$ NH$_3$/HC$_3$N pairs observed at lower 
spectral resolution (FWHM = 120--500 \ms) in the IRDCs inferred that
$\Delta V = 160\pm32_{\rm stat}\pm14_{\rm sys}$ \ms\ and
$120\pm37_{\rm stat}\pm31_{\rm sys}$ \ms, respectively.
At the same time,
the rotational--rotational velocity differences for the IRDCs did not show any significant
offset: 
$\Delta V = V$(\nnhp) -- $V$(\hcccn) = $-17\pm34$ \kms. 

The physical properties of low-mass ($M \la 10 M_\odot$) molecular cores observed with the GBT 
are similar to those analyzed in the present paper. 
However, the IRDCs are more massive
objects ($M \ga 100 M_\odot$) with high turbulent velocities ($\sim$1-3 \kms). 
The IRDCs are also more distant clouds, which are located towards the Galactic center at distances
of 2-5 kpc. 
The spectral resolution used in observations of the IRDCs does not
allow us to resolve the
hyperfine structure of the HC$_3$N $J=5-4$ transition (Sakai \etal\ 2008).
The spectra of the HC$_3$N lines were fitted to a single
Gaussian function and the corresponding radial velocities
were calculated based on a single
hyperfine line $J=5-4, F=5-4$
with the frequency 45490.3102(3) MHz taken from the JPL catalog\footnote{http://spec.jpl.nasa.gov}.
As a result, a systematic shift related to unresolved hyperfine components
was introduced. 
We also note that the JPL frequencies are systematically shifted
with respect to the CDMS catalogue (M\"uller \etal\ 2005).
The CDMS frequency of the $J=5-4, F=5-4$ transition is, for example,
45490.3137(5) MHz. 
Thus, it seems plausible that the large offset, $\Delta V \ga 100$ \ms, 
deduced from the IRDCs is caused by the superposition of unaccounted systematic shifts.

As for the low-mass cores in the Perseus molecular cloud and the Pipe Nebula, 
the discrepancy between our present estimate,
$\Delta V \simeq 20$ \ms, and the previously obtained value,
$\Delta V \simeq 60$ \ms, is probably caused by uncertainties in the rest-frame frequency
of the CCS $J_N = 2_1-1_0$ transition. 
In particular, the frequency 22344.033(1) MHz of this transition
(used by both Rosolowsky \etal\ and Rathborne \etal)
was calculated by comparing with the HC$_3$N $J = 5-4$ line
observed towards a cold dark cloud L1498 assuming that ($i$) CCS and HC$_3$N
are co-spatially distributed 
and ($ii$) the rest-frame frequency of HC$_3$N $J = 5-4$, $F = 5-4$ is 45490.316(1) MHz
(Yamamoto \etal\ 1990). If the HC$_3$N frequency is 2 kHz lower as indicated in the
CDMS catalogue, then the laboratory frequency of CCS decreases by 1 kHz, where
1 kHz at 22.3 GHz corresponds to 13.4 \ms, i.e., the difference
$V$(CCS) -- $V$(\amm) becomes equal to $\simeq 46$ \ms.
The only known laboratory measurement of the CCS frequency
infers, however, that 22344.029(4) MHz (Lovas \etal\ 1992). With this value 
$V$(CCS) -- $V$(\amm) $\simeq$ 6 \ms, but its error reaches 53 \ms\ (4 kHz at 22.3 GHz). 
The rest-frame frequency of CCS reported in the CDMS is 22344.0308(10) MHz. With this frequency, 
we obtain for the Perseus molecular cloud the offset
$\Delta V = 22\pm7_{\rm stat}\pm14_{\rm sys}$ \ms, which is 
in full agreement with our present estimate,
$\Delta V = 23\pm4_{\rm stat}\pm3_{\rm sys}$ \ms. 
This consideration makes it quite obvious that
to increase the reliability of the results obtained, high precision laboratory measurements
of molecular transition frequencies are badly needed.

\section{Conclusions and future prospects}
\label{sect-7}

Using the 32-m Medicina, 100-m Effelsberg, and 45-m Nobeyama radio telescopes, we 
have performed precise spectral measurements of the relative radial velocity differences between
the rotational transitions in \hcccn\ $(J=2-1)$ and \nnhp\ ($J=1-0$) and the inversion
transition in \amm\ $(J,K)=(1,1)$. 
We detect a velocity offset of
$\Delta V = 23\pm4_{\rm stat}\pm3_{\rm sys}$ \ms\ between the rotational and inversion
lines observed in cold molecular cores with dominating thermal motion. 
We do not find any plausible systematic effects that
could mimic an offset of about 20 \ms\ between rotational and inversion transitions
and would be regularly reproduced in observations of different cold
molecular cores with different facilities. 
The measured positive offset is qualitatively consistent with our preliminary result,
$\Delta V \simeq 60$ \ms, obtained on the basis of the GBT 
observations of Rosolowsky \etal\ (2008) and Rathborne \etal\ (2008) 
of the NH$_3$/CCS pairs in
molecular cores in the Perseus molecular cloud and the Pipe Nebula (LMK). However,
the rest-frame frequency of the CCS $J_N = 2_1-1_0$ transition is not well known, and before
deciding whether the difference between the current and preliminary results is statistically
significant or not, new high precision laboratory measurements of the CCS frequency
should be carried out.

If we assume that the measured velocity offset is caused by the electron-to-proton
mass ratio variation, then  
\dmm = $22\pm4_{\rm stat}\pm3_{\rm sys}$ ppb.
To account for the conditions of the terrestrial 
laboratory experiments, the non-zero $\Delta\mu$
would require chameleon-like scalar field models that predict a strong
dependence of mass and coupling constant on the ambient matter density.

In any case, our estimate defines a strongest conservative upper limit to
$|\Delta\mu/\mu| \leq 3\times10^{-8}$, which can be considered as a reference point
at $z = 0$.
We note that extragalactic molecular clouds have 
gas densities similar to those in the interstellar clouds of the Milky Way. 
Thus, the value of 
\dmm\ in high-$z$  molecular systems is expected to be at the same level as
in the interstellar clouds, i.e., \dmm $\sim 10^{-8}$, provided that 
no temporal dependence of the electron-to-proton mass ratio is present.

To be completely confident
that the derived velocity shift is not caused by kinematic effects in the
clouds but reflects the density-modulated  variation of \dmm,
new high precision radio-astronomical
observations are needed for a wider range of objects. These
observations should also target different molecules (i.e., not NH$_3$) 
with tunneling transitions sensitive to the changes in $\mu$ and
molecules with $\Lambda$-doublet lines that also exhibit enhanced
sensitivity to variations in $\mu$ and $\alpha$ (Kozlov  2009).

For the molecular transitions in question, it is very important to 
measure their rest-frame 
frequencies with an accuracy of about 1 \ms\ in laboratory experiments.
In some cases the present uncertainties in the rest-frame frequencies are larger
than the errors of radio-astronomical measurements, thus preventing 
unambiguous conclusions. 

In addition, searches for variations in the fine-structure
constant $\alpha$ within the Milky Way disk using mid- and
far-infrared fine-structure transitions in atoms and ions (Kozlov \etal\ 2008),
or searches for variations in the combination of $\alpha^2/\mu$
using the [\ion{C}{ii}] $\lambda158$ $\mu$m line and
CO rotational lines (Levshakov \etal\ 2008b), or the [\ion{C}{i}] $\lambda609$ $\mu$m line and
low-lying rotational lines of $^{13}$CO (Levshakov \etal\ 2009)
would be of great importance for cross-checking these results.

\begin{table*}[t!]
\centering
\caption{Target list. 
}
\label{tbl-1}
\begin{tabular}{rlcccllc}
\hline
\hline
\noalign{\smallskip}
 & \multicolumn{1}{c}{Source} & R.A. & Dec.  & $V_{lsr}$ & 
\multicolumn{1}{c}{Obs.$^a$} & \multicolumn{1}{c}{Molecules$^b$} & References$^c$\\
\multicolumn{1}{c}{\#} & & \multicolumn{1}{c}{(J2000.0)} & 
\multicolumn{1}{c}{(J2000.0)} & (\kms) \\
\noalign{\smallskip}
\hline 
\noalign{\smallskip}
1 &L1355 & 02:53:12.1 & $+$68:55:52 & $-3.7$ & $m, e$   & \amm\  & 14\\ 
2 &NH3SRC8 & 03:25:26.3 & $+$30:45:05 & $+4.1$ & $m, e, n $ & \amm, \nnhp  & 5\\
3 &NH3SRC28 & 03:27:26.4 & $+$29:51:08 & $+5.6$ & $e$   & \amm\  & 5\\
4 &NH3SRC36 & 03:27:55.9 & $+$30:06:18 & $+4.7$ & $m$   & \amm\ & 5\\
5 &NH3SRC91 & 03:30:15.2 & $+$30:23:39 & $+5.9$ & $m$ &\amm & 5\\
6 &NH3SRC95 & 03:30:32.1 & $+$30:26:19 & $+6.1$ & $m$   & \amm, \hcccn & 5\\
7 &NH3SRC97 & 03:30:50.5 & $+$30:49:17 & $+7.7$ & $e$   & \amm & 5\\
8 &L1489    & 04:04:49.0 & $+$26:18:42 & $+6.7$ & $e, n$ & \amm,  \hcccn, \nnhp & 1, 4\\ 
9 &L1498    & 04:10:51.4 & $+$25:09:58 & $+7.8$ & $e, n$ & \amm, \hcccn, \nnhp & 1, 3, 4, 6\\
10 &L1495    & 04:14:08.2 & $+$28:08:16 & $+6.8$ & $e, n$ & \amm, \hcccn, \nnhp & 1, 6\\
11 &L1521F   & 04:28:39.8 & $+$26:51:35 & $+6.5$ & $e$ & \amm, \hcccn & 4, 6, 10\\ 
12 &L1524    & 04:29:22.7 & $+$24:34:58 & $+6.5$ & $e, n$ & \amm, \hcccn, \nnhp & 4, 8, 10\\ 
13 &B217S-W   & 04:27:46.4 & $+$26:17:52 & $+7.0$ & $e$ & \amm, \hcccn & 1, 4, 12\\ 
14 &L1400K    & 04:30:52.0 & $+$54:51:55 & $+3.3$ & $e, n$ & \amm, \hcccn & 1, 3, 4, 6\\ 
15 &TMC-2     & 04:32:48.6 & $+$24:25:12 & $+6.2$ & $e, n$ & \amm, \hcccn, \nnhp & 1\\
16 &L1536     & 04:33:23.5 & $+$22:42:46 & $+5.6$ & $e, n$ & \amm, \hcccn, \nnhp & 1, 4\\ 
17 &CB22      & 04:40:33.1 & $+$29:55:04 & $+6.0$ & $e$ & \amm, \hcccn & 4\\ 
18 &TMC-1C    & 04:41:38.8 & $+$26:00:42 & $+5.2$ & $m,e,n$ & \amm, \hcccn, \nnhp & 1, 2, 3, 4\\
19 &TMC-1/HC3N& 04:41:41.8 & $+$25:41:42 & $+5.9$ & $m$ & \amm, \hcccn & 9, 11\\ 
20 &CB23      & 04:43:27.7 & $+$29:39:11 & $+6.0$ & $e, n$ & \amm, \hcccn, \nnhp & 4, 6\\
21 &L1517B    & 04:55:18.8 & $+$30:38:04 & $+5.8$ & $e, n$ & \amm, \hcccn, \nnhp & 2, 4, 6\\
22 &L1512     & 05:04:09.6 & $+$32:43:09 & $+7.1$ & $m, e, n$ & \amm, \hcccn, \nnhp & 1, 2, 4, 6, 8\\ 
23 &L1544     & 05:04:16.5 & $+$25:10:48 & $+7.2$ & $e, n$ & \amm, \hcccn & 3, 4, 6, 8, 9\\
24 &L134A     & 15:53:36.2 & $-$04:35:26 & $+2.7$ & $e$ & \amm & 4, 6, 8\\ 
25 &L183      & 15:54:06.4 & $-$02:52:23 & $+2.4$ & $m, e, n$ & \amm, \hcccn, \nnhp & 4, 6, 7, 8\\ 
26 &L43       & 16:34:35.0 & $-$15:46:36 & $+0.7$ & $e, n$ & \amm, \hcccn, \nnhp & 4 \\ 
27 &L260-NH3  & 16:47:06.6 & $-09$:35:21 & $+3.5$ & $e, n$ & \amm, \hcccn, \nnhp & 1, 8, 12\\ 
28 &L234A     & 16:48:06.8 & $-$10:51:48 & $+2.9$ & $e$ & \amm, \hcccn & 4, 8\\ 
29 &L234S-E   & 16:48:08.6 & $-$10:57:25 & $+3.2$ & $e$ & \amm & 4\\
30 &L63       & 16:50:15.4 & $-$18:06:06 & $+5.7$ & $e$ & \amm, \hcccn & 4, 8\\ 
31 &CB68      & 16:57:19.4 & $-$16:09:25 & $+5.2$ & $e$ & \amm, \hcccn & 4 \\ 
32 &L492      & 18:15:49.6 & $-$03:46:14 & $+7.7$ & $e$ & \amm, \hcccn & 6\\ 
33 &L778      & 19:26:32.5 & $+$23:58:42 & $+9.9$ & $m$ & \amm & 1, 4\\ 
34 &B335      & 19:37:01.3 & $+$07:34:07 & $+8.4$ & $e$ & \amm, \hcccn & 4, 8\\ 
35 &L694-2    & 19:41:04.4 & $+$10:57:02 & $+9.5$ & $m, e$ & \amm & 6, 13\\ 
36 &L1049-2   & 20:41:30.5 & $+$57:33:44 & $+0.1$ & $e$ & \amm & 15 \\ 
37 &L1251T3   & 22:29:50.3 & $+$75:13:25 & $-3.8$ & $e, n$ & \amm, \hcccn, \nnhp & 4 \\ 
38 &L1251T2   & 22:30:31.2 & $+$75:13:38 & $-4.0$ & $e, n$ & \amm, \hcccn, \nnhp & 4 \\ 
39 &L1251T1   & 22:31:02.3 & $+$75:13:39 & $-4.2$ & $e, n$ & \amm, \hcccn, \nnhp & 4 \\
40 &L1251C    & 22:35:53.6 & $+$75:18:55 & $-4.7$ & $e, n$ & \amm, \hcccn, \nnhp & 16,17 \\ 
41 &L1197     & 22:37:02.1 & $+$58:57:22 & $-3.1$ & $e, n$ & \amm, \hcccn, \nnhp & 6,13 \\ 
\noalign{\smallskip}
\hline
\noalign{\smallskip}
\multicolumn{8}{l}{$^a$ Observations: $m$~-- Medicina 32-m, $e$~-- Effelsberg 100-m, 
and $n$~-- Nobeyama 45-m radio telescopes. }\\
\multicolumn{8}{l}{$^b$ Detected transitions: 
NH$_3$ $(J,K) = (1,1)$,  HC$_3$N $J = 2-1$, N$_2$H$^+$ $J = 1-0$.  }\\
\multicolumn{8}{l}{$^c$ References: (1) Benson \& Myers 1989; (2) Myers \etal\ 1983; 
(3) Howe \etal\ 1994; (4) Jijina \etal\ 1999;  }\\
\multicolumn{8}{l}{\,\, (5) Rosolowsky \etal\ 2008; (6) Crapsi \etal\ 2005; (7) Pagani \etal\ 2009;
(8) Fuller \& Myers 1993;  }\\
\multicolumn{8}{l}{\,\, (9) Churchwell \etal\ 1984; (10) Codella \etal\ 1997; 
(11) Benson \& Myers 1983; (12) Motte \& Andre 2001;}\\
\multicolumn{8}{l}{\,\, (13) Lee \etal\ 2007; (14) Park \etal\ 2004; (15) 
Lee \etal\ 1999; (16) Caselli  \etal\ 2002; (17) Walsh \etal\ 2004.}\\

\end{tabular}
\end{table*}

\begin{table}[t!]
\centering
\caption{Hyperfine components of the NH$_3$ $(J,K) = (1,1)$ transition.   
The velocity offsets, $v_i$, are relative to 23.694495487 GHz. Data from
Kukolich (1967) and Ho (1977). 
Shown in parentheses are $1\sigma$ errors.
}
\label{tbl-2}
\begin{tabular}{ccccccr@{.}l}
\hline
\hline
\noalign{\smallskip}
\# & $F'_1$ & $F'$ & $F_1$ & $F$ & Weight & \multicolumn{2}{c}{$v_i$, \ms} \\ 
\noalign{\smallskip}
\hline
\noalign{\medskip}
1& 1 & 1.5 & 0 & 0.5 & 0.317 & $-19549$&98(48)  \\[-2pt]
 & 1 & 0.5 & 0 & 0.5 & 0.159 & $-19411$&73(61)  \\[2pt]
2& 1 & 1.5 & 2 & 2.5 & 0.357 & $-7815$&25(63)  \\[-2pt]
 & 1 & 1.5 & 2 & 1.5 & 0.040 & $-7372$&80(75)  \\[-2pt]
 & 1 & 0.5 & 2 & 1.5 & 0.198 & $-7233$&48(58)  \\[2pt]
3 & 2 & 1.5 & 2 & 2.5 & 0.071 & $-250$&92(60)  \\[-2pt]
 & 1 & 1.5 & 1 & 1.5 & 0.198 & $-213$&00(47)  \\[-2pt]
 & 2 & 2.5 & 2 & 2.5 & 1.000 & $-132$&38(47)  \\[-2pt]
 & 1 & 0.5 & 1 & 1.5 & 0.040 & $-75$&17(75)  \\[-2pt]
 & 2 & 1.5 & 2 & 1.5 & 0.643 & $192$&27(56)  \\[-2pt]
 & 2 & 2.5 & 2 & 1.5 & 0.071 & $311$&03(80)  \\[-2pt]
 & 1 & 1.5 & 1 & 0.5 & 0.040 & $322$&04(35)  \\[-2pt]
 & 1 & 0.5 & 1 & 0.5 & 0.079 & $460$&41(46)  \\[2pt]
4 & 2 & 1.5 & 1 & 1.5 & 0.040 & $7351$&32(49)  \\[-2pt]
 & 2 & 2.5 & 1 & 1.5 & 0.357 & $7469$&67(73)  \\[-2pt]
 & 2 & 1.5 & 1 & 0.5 & 0.198 & $7886$&69(72)  \\[2pt]
5 & 0 & 0.5 & 1 & 1.5 & 0.317 & $19315$&90(71)  \\[-2pt]
 & 0 & 0.5 & 1 & 0.5 & 0.159 & $19851$&27(35)$^\dagger$  \\[-2pt]
\noalign{\smallskip}
\hline
\noalign{\smallskip}
\multicolumn{8}{l}{$^\dagger$The error of 0.35 \ms\ corresponds to the difference}\\[-2pt]
\multicolumn{8}{l}{between theoretically predicted (Ho 1977) and astrono-}\\[-2pt] 
\multicolumn{8}{l}{mically determined (Rydbeck \etal\ 1977) frequencies.}
\end{tabular}
\end{table}

\begin{table}[t!]
\centering
\caption{Hyperfine components of the HC$_3$N $J = 2-1$ transition.
The velocity offsets, $v_i$, are relative to 18.19621694  GHz. 
Data from Lapinov (2008, private comm.) and M\"uller \etal\ (2005). 
}
\label{tbl-3}
\begin{tabular}{ccccr}
\hline
\hline
\noalign{\smallskip}
\# & $F'$ & $F$ & Weight & \multicolumn{1}{c}{$v_i$, \ms} \\ 
\noalign{\smallskip}
\hline
\noalign{\medskip}
1 & 1 & 1 &  0.333 &   $-35548$.77(2.82)\\[-2pt]  
2 & 1 & 2 &  0.022  &  $-14167$.97(2.83)\\[-2pt] 
3 & 3 & 2 &  1.867  &  $-1540$.96(2.82)\\[-2pt] 
4 & 2 & 1 &  1.000 &      0.00(2.82)\\[-2pt]    
5 & 1 & 0 &  0.444  &   17806.60(2.82)\\[-2pt]
6 & 2 & 2 &  0.333  &   21380.64(2.82)\\ 
\noalign{\smallskip}
\hline
\end{tabular}
\end{table}

\begin{table}[t!]
\centering
\caption{Hyperfine components of the N$_2$H$^+$ $J = 1-0$ transition.   
The velocity offsets, $v_i$, are relative to 93.1737722 GHz. 
Data from M\"uller \etal\ (2005).
}
\label{tbl-4}
\begin{tabular}{ccccccr}
\hline
\hline
\noalign{\smallskip}
\# & $F'_1$ & $F'$ & $F_1$ & $F$ & Weight & \multicolumn{1}{c}{$v_i$, \ms} \\ 
\noalign{\smallskip}
\hline
\noalign{\medskip}
1 &0&1 &1&0& 0.076 & $-8008.2(13.8)$  \\[-2pt]
 &0&1 &1&1& 0.100 & $-8008.2(13.8)$  \\[-2pt]
 &0&1 &1&2& 0.252 & $-8008.2(13.8)$  \\[2pt]
2&2&1 &1&2& 0.017 & $-622.6(13.8)$  \\[-2pt]
 &2&1 &1&0& 0.134 & $-622.6(13.8)$  \\[-2pt]
&2&1 &1&1& 0.278 & $-622.6(13.8)$  \\[2pt]
3&2&3 &1&2& 1.000 & 0.00(13.2)  \\[2pt]
4&2&2 &1&2& 0.111 & 956.9(13.5)  \\[-2pt]
 &2&2 &1&1& 0.604 & 956.9(13.5)  \\[2pt]
5&1&1 &1&1& 0.051 & 5541.6(13.5)  \\[-2pt]
 &1&1 &1&2& 0.159 & 5541.6(13.5)  \\[-2pt]
 &1&1 &1&0& 0.218 & 5541.6(13.5)  \\[2pt]
6&1&2 &1&1& 0.111 & 5982.4(13.5) \\[-2pt]
 &1&2 &1&2& 0.604 & 5982.4(13.5)  \\[2pt]
7&1&0 &1&1& 0.143 & 6934.2(13.2)  \\
\noalign{\smallskip}
\hline
\end{tabular}
\end{table}

\begin{table*}[t!]
\centering
\caption{Radial velocities, $V_{lsr}$, Doppler parameters, $b$, and corresponding 
$\chi^2_\nu$ values normalized per degree of freedom. Data from the Medicina 
32-m and Effelsberg 100-m radio telescopes.
The numbers in parentheses correspond to $1\sigma$ statistical errors.
}
\label{tbl-5}
\begin{tabular}{l r@{.}l r@{.}l  r@{.}l r@{.}l r@{.}l r@{.}l r@{.}l  c}
\hline
\hline
\noalign{\smallskip}
 & \multicolumn{14}{c}{ $V_{lsr}$, \kms\,/\,b, \kms\,/\,$\chi^2_\nu$ } \\[2pt] 
 & \multicolumn{8}{c}{ NH$_3$ (1,1) } & \multicolumn{6}{c}{ HC$_3$N $(2-1)$ } & $\Delta V$  \\[2pt]
\multicolumn{1}{c}{Object} &  \multicolumn{2}{c}{outer} & \multicolumn{2}{c}{inner} & 
\multicolumn{2}{c}{central} & 
\multicolumn{2}{c}{total} & \multicolumn{2}{c}{low} & \multicolumn{2}{c}{high} &
\multicolumn{2}{c}{total} & \multicolumn{1}{c}{(\ms)} \\
\noalign{\smallskip}
\hline 
\noalign{\smallskip}

\multicolumn{16}{c}{\it Medicina 32-m radio telescope} \\
\noalign{\smallskip}

L1512 &  7&0800(60) & 7&0795(47) & 7&0760(41) & 7&0790(28) & \multicolumn{2}{c}{ } & 7&0965(30) & 
\multicolumn{2}{c}{ } & +17.5(4.1) \\[-2pt]
&  0&1299(93) & 0&116(18) & 0&1138(51) & 0&1189(47) & \multicolumn{2}{c}{ } &  0&1037(96) & \multicolumn{2}{c}{ } \\[-2pt]
&  1&04 & 0&89 & 0&38 & 0&89  & \multicolumn{2}{c}{ } & 0&78 & \multicolumn{2}{c}{ } \\
L183 &  2&4000(63) & 2&3970(49) & 2&4140(60) & 2&4075(32) & \multicolumn{2}{c}{ } & 2&404(15) & 
\multicolumn{2}{c}{ } & $-3.5(15.3)$ \\[-2pt]
&  0&129(18) & 0&148(13) & 0&1297(55) & 0&1363(51) & \multicolumn{2}{c}{ } &  0&134(61) & \multicolumn{2}{c}{ } \\[-2pt]
&  1&16 & 0&88 & 0&74 & 1&30  & \multicolumn{2}{c}{ } & 0&97 & \multicolumn{2}{c}{ } \\

\multicolumn{16}{c}{\it Effelsberg 100-m radio telescope} \\
\noalign{\smallskip}

L1498 &  7&8025(16) & 7&8080(13) & 7&8065(13) & 7&8060(8) & 7&8290(38) & 7&8280(15) & 7&8280(14) & +22.0(1.6) \\[-2pt]
&  0&1125(31) & 0&1138(64) & 0&1227(14) & 0&1177(13)  & 0&105(13) & 0&0904(52) & 0&0910(48)  \\[-2pt]
&  1&64 & 0&87 & 0&80 & 1&22  & 1&19 & 0&55 & 0&99 \\
L1495 &  6&7690(30) & 6&7830(27) & 6&7895(19) & 6&7840(14) & \multicolumn{2}{c}{ } & 6&8175(38) & 6&8175(37) & +33.5(4.0) \\[-2pt]
&  0&1421(57) & 0&147(17) & 0&1425(20) & 0&1446(19)  & \multicolumn{2}{c}{ } & 0&113(14) & 0&115(12)  \\[-2pt]
&  1&36 & 0&69 & 0&87 & 1&09  & \multicolumn{2}{c}{ } & 0&86 & 0&89 \\
L1400K &  3&2580(28) & 3&2555(21) & 3&2650(19) & 3&2600(13) & 3&2690(50) & 3&2780(20) & 3&2770(18) & +17.0(2.2) \\[-2pt]
&  0&1277(50) & 0&1111(59) & 0&1224(24) & 0&1165(20)  & 0&107(16) & 0&0986(96) & 0&0961(94)  \\[-2pt]
&  1&40 & 1&00 & 0&88 & 1&18  & 0&52 & 0&70 & 1&0 \\
CB22 &  5&9665(26) & 5&9675(20) & 5&9740(17) & 5&9705(11) & \multicolumn{2}{c}{ } & 5&9945(30) & 5&9935(28) & +23.0(3.0) \\[-2pt]
&  0&1256(60) & 0&125(11) & 0&1357(19) & 0&1305(18)  & \multicolumn{2}{c}{ } & 0&088(17) & 0&086(14)  \\[-2pt]
&  1&10 & 0&89 & 0&75 & 1&02  & \multicolumn{2}{c}{ } & 0&44 & 1&02 \\
TMC-1C$^a$ &  5&2405(36) & 5&2570(36) & 5&2575(28) & 5&2540(19) & 5&2815(56) & 5&2880(22) & 5&2875(20) & +33.5(2.8) \\[-2pt]
&  0&0997(70) & 0&126(10) & 0&1135(32) & 0&1213(29)  & 0&086(24) & 0&1055(70) & 0&1032(66)  \\[-2pt]
&  1&17 & 0&85 & 0&43 & 0&94  & 0&75 & 1&11 & 0&92 \\
L1517B &  5&7800(20) & 5&7830(17) & 5&7855(18) & 5&7835(10) & 5&8200(76) & 5&8250(23) & 5&8245(20) & +39.0(1.5) \\[-2pt]
&  0&1131(40) & 0&1229(61) & 0&1201(17) & 0&1216(16)  & 0&126(60) & 0&1073(69) & 0&1100(68)  \\[-2pt]
&  1&18 & 0&79 & 0&55 & 0&96  & 0&91 & 0&70 & 0&82 \\
L1512 &  7&1165(24) & 7&1120(13) & 7&1130(13) & 7&1130(8) & \multicolumn{2}{c}{ } & 7&1345(9) & 7&1335(9) & +20.5(1.2) \\[-2pt]
&  0&1244(35) & 0&1166(62) & 0&1135(15) & 0&1132(13)  & \multicolumn{2}{c}{ }  & 0&0784(43) & 0&0786(41)  \\[-2pt]
&  1&30 & 0&76 & 0&77 & 1&03  & \multicolumn{2}{c}{ } & 0&94 & 1&04 \\
L183 &  2&4200(13) & 2&4170(11) & 2&4220(12) & 2&4195(7) & \multicolumn{2}{c}{ } & 2&4335(32) & 2&4320(30) & +12.5(3.1) \\[-2pt]
&  0&1062(25) & 0&1149(40) & 0&1166(13) & 0&1118(11)  & \multicolumn{2}{c}{ }  & 0&097(18) & 0&104(11)  \\[-2pt]
&  1&14 & 0&75 & 0&54 & 0&98  & \multicolumn{2}{c}{ } & 0&71 & 1&00 \\
L260-NH$_3$ &  3&4605(15) & 3&4635(11) & 3&4580(15) & 3&4615(8) & 3&4565(50) & 3&4645(21) & 3&4630(20) & +1.5(2.2) \\[-2pt]
&  0&1089(30) & 0&1082(39) & 0&1126(15) & 0&1084(13)  & 0&084(14)  & 0&0973(72) & 0&0951(61)  \\[-2pt]
&  1&58 & 0&93 & 1&16 & 1&39  & 1&26 & 0&84 & 1&15 \\
L234A &  2&8935(50) & 2&9065(57) & 2&9030(35) & 2&9030(25) & \multicolumn{2}{c}{ } & 2&9165(40) & 2&9185(45) & +15.5(5.1) \\[-2pt]
&  0&1337(93) & 0&170(45) & 0&1601(33) & 0&1596(32)  & \multicolumn{2}{c}{ }  & 0&136(14) & 0&135(15)  \\[-2pt]
&  0&98 & 0&73 & 0&83 & 0&90  & \multicolumn{2}{c}{ } & 0&92 & 0&73 \\
B335 &  8&3350(25) & 8&3430(27) & 8&3250(20) & 8&3365(14) & \multicolumn{2}{c}{ } & 8&3545(51) & 8&3585(50) & +22.0(5.2) \\[-2pt]
&  0&2312(55) & 0&2326(36) & 0&2138(22) & 0&2294(19)  & \multicolumn{2}{c}{ }  & 0&196(12) & 0&195(13)  \\[-2pt]
&  0&84 & 1&29 & 0&89 & 1&14  & \multicolumn{2}{c}{ } & 0&61 & 0&99 \\
L1251C &  $-4$&7390(22) & $-4$&7360(21) & $-4$&7200(21) & $-4$&7320(12) & \multicolumn{2}{c}{ } & $-4$&7005(70) & $-4$&6975(70) 
& +34.5(7.1) \\[-2pt]
&  0&1396(55) & 0&1614(64) & 0&1577(21) & 0&1588(18)  & \multicolumn{2}{c}{ }  & 0&146(18) & 0&141(19)  \\[-2pt]
&  1&77 & 1&18 & 0&67 & 1&48  & \multicolumn{2}{c}{ } & 0&65 & 0&74 \\
\noalign{\smallskip}
\hline
\noalign{\smallskip}
\multicolumn{16}{l}{$^a$ Offset $(\Delta\alpha,\Delta\delta) = (0'',40'')$. } 
\end{tabular}
\end{table*}

\begin{table*}[t!]
\centering
\caption{Radial velocities, $V_{lsr}$, Doppler parameters, $b$, and corresponding 
$\chi^2_\nu$ values normalized per degree of freedom. Data from the Nobeyama 45-m radio telescope.
The numbers in parentheses correspond to $1\sigma$ statistical errors.
}
\label{tbl-6}
\begin{tabular}{l r@{.}l r@{.}l  r@{.}l r@{.}l r@{.}l r@{.}l r@{.}l  c}
\hline
\hline
\noalign{\smallskip}
 & \multicolumn{14}{c}{ $V_{lsr}$, \kms\,/\,b, \kms\,/\,$\chi^2_\nu$ } \\[2pt] 
 & \multicolumn{8}{c}{ NH$_3$ (1,1) } & \multicolumn{6}{c}{ N$_2$H$^+$ $(1-0)$ } & $\Delta V$  \\[2pt]
\multicolumn{1}{c}{Object} &  \multicolumn{2}{c}{outer} & \multicolumn{2}{c}{inner} & \multicolumn{2}{c}{central} & 
\multicolumn{2}{c}{total} & \multicolumn{2}{c}{low} & \multicolumn{2}{c}{high} &
\multicolumn{2}{c}{total} & \multicolumn{1}{c}{(\ms)} \\
\noalign{\smallskip}
\hline 
\noalign{\smallskip}
L1498 &  7&8215(30) & 7&8210(25) & 7&8180(25) & 7&8200(15) & 7&8450(42) & 7&8400(43) & 7&8430(30) & +23.0(3.4) \\[-2pt]
&  0&1080(54) & 0&116(11) & 0&1129(30) & 0&1124(26) & 0&092(15) &  0&0987(40) & 0&1025(68) \\[-2pt]
&  1&37 & 0&65 & 1&07 & 1&00  & 0&84 & 1&16 & 1&08 \\
L1536 &  5&5900(28) & 5&5935(24) & 5&6030(27) & 5&5955(15) & 5&629(11) & 5&6295(92) & 5&6290(80) & +33.5(8.1) \\[-2pt]
&  0&1284(66) & 0&1306(94) & 0&1391(27) & 0&1315(26) & 0&121(24) &  0&117(40) & 0&118(20) \\[-2pt]
&  0&90 & 1&16 & 1&07 & 1&15  & 0&76 & 0&77 & 0&75  \\
CB23 &  6&0345(30) & 6&0365(22) & 6&0325(22) & 6&0345(14) & 6&0515(46) & 6&0385(45) & 6&0460(33) & +11.5(3.6) \\[-2pt]
&  0&1027(65) & 0&106(11) & 0&1071(28) & 0&1064(25) & 0&097(13) &  0&119(12) & 0&1043(92) \\[-2pt]
&  1&47 & 0&84 & 0&86 & 1&08  & 0&64 & 1&40 & 1&00 \\
L1517B &  5&8015(62) & 5&7980(60) & 5&7965(58) & 5&7985(34) & 5&8190(50) & 5&8215(53) & 5&8195(38) & +21.0(5.1) \\[-2pt]
&  0&107(12) & 0&119(28) & 0&1228(64) & 0&1171(57) & 0&109(21) &  0&138(16) & 0&127(14) \\[-2pt]
&  1&13 & 0&45 & 1&15 & 0&84  & 1&00 & 1&03 & 1&03 \\
L1512 &  7&0925(59) & 7&1030(38) & 7&1080(42) & 7&1025(25) & 7&1290(45) & 7&1260(41) & 7&1270(35) & +24.5(4.3) \\[-2pt]
&  0&127(14) & 0&112(21) & 0&1210(48) & 0&1158(43) & 0&0847(92) &  0&085(11) & 0&0850(90) \\[-2pt]
&  0&59 & 0&93 & 1&00 & 0&94  & 0&89 & 0&89 & 0&86 \\
L183 &  2&4195(27) & 2&4165(26) & 2&4275(33) & 2&4225(16) & 2&4395(50) & 2&4485(45) & 2&4445(32) & +22.0(3.6) \\[-2pt]
&  0&1490(78) & 0&1563(65) & 0&1536(29) & 0&1563(27) & 0&149(17) &  0&1484(61) & 0&154(56) \\[-2pt]
&  1&55 & 1&01 & 1&75 & 1&64  & 0&91 & 0&82 & 0&91 \\
L183$^a$ &  2&4245(19) & 2&4265(18) & 2&4375(27) & 2&4265(11) & 2&4190(26) & 2&4135(26) & 2&4165(18) 
& $-10.0(2.1)$ \\[-2pt]
&  0&1579(50) & 0&1681(45) & 0&1816(22) & 0&1610(20) & 0&1325(64) &  0&1624(69) & 0&1423(54) \\[-2pt]
&  1&18 & 0&82 & 1&32 & 1&43  & 1&18 & 1&01 & 1&30 \\
L43 & 0&7020(25) & 0&7040(27) & 0&7150(27) & 0&7050(16) & 0&7485(31) & 0&7470(25) & 0&7480(19) & +43.0(2.5) \\[-2pt]
&  0&1523(63) & 0&1748(64) & 0&1859(29) & 0&1677(25) & 0&1368(88) &  0&173(17) & 0&1512(93) \\[-2pt]
&  1&46 & 1&37 & 0&84 & 1&43  & 1&81 & 0&90 & 1&60 \\
L260-NH$_3$ & 3&4815(25) & 3&4800(21) & 3&4750(25) & 3&4790(14) & 3&4700(59) & 3&4710(52) & 3&4690(37) 
& $-10.0(4.0)$ \\[-2pt]
&  0&1053(55) & 0&119(10) & 0&1144(26) & 0&1131(23) & 0&0997(82) &  0&111(26) & 0&1061(79) \\[-2pt]
&  1&26 & 0&83 & 0&53 & 1&06  & 1&34 & 1&12 & 1&24 \\
\noalign{\smallskip}
\hline
\noalign{\smallskip}
\multicolumn{16}{l}{$^a$ Offset $(\Delta\alpha,\Delta\delta) = (-31.5'',25'')$. } 
\end{tabular}
\end{table*}

\begin{acknowledgements}
We are grateful to the staffs of the Medicina,
Effelsberg, and Nobeyama radio observatories for excellent supports in
our observations.
The authors thank Dieter Engels for assistance in observations at the
Medicina 32-m telescope. 
The project has been supported by
DFG Sonderforschungsbereich SFB 676 Teilprojekt C4
and by the RFBR grant No. 09-02-12223.
S.A.L. is supported by the RFBR grant No. 09-02-00352-a, 
and by the Federal Agency for Science and Innovations grant
NSh 2600.2008.2.
A.V.L. is supported by the RFBR grant No. 08-02-92001 and
by the Program IV.12/2.5 of the Physical Department of the RAS.
\end{acknowledgements}

\begin{appendix}
\section{Robust statistics}
\label{AA}

A comprehensive review of robust statistical procedures is given 
by Hampel \etal\ (1986) (see also Chap.~15.7 in Press \etal\ 1992).

Robust estimates are insensitive to departures of the distribution shape from the assumed data
distribution. In particular, $M$-estimates (maximum-likelihood)
are usually relevant for the estimate of parameters with continuous
distributions. The shift (mean), $m$, and the scale (standard deviation),
$s$, of a sample of data points $\{x_i\}$ are estimated by
minimizing the sum $\sum_i \Psi[(x_i - m)/s]$.
Here $\Psi$ is a weighting function such that the weights given to
individual points first increase with deviations from $m$, and then
decrease so that very outlying points (outliers) are not counted at all
(re-descending $M$-estimate). The following functions are prescribed for
use as $\Psi$:

\noindent
Hampel's function
\[ \Psi(u) = \mbox{sign}(u) \left\{
\begin{array}{ll}
|u|, & 0 \leq |u| \leq 1.7 \\
1.7, & 1.7 \leq |u| \leq 3.4 \\
\frac{8.5-|u|}{3}, & 3.4 \leq |u| \leq 8.5 \\
0, & u > 8.5
\end{array} \right. \]

\noindent
Andrew's sine
\[ \Psi(u) = \left\{
\begin{array}{ll}
\mbox{sin}\left(\frac{u}{c}\right), & |u| < c\pi \\
0, & |u| > c\pi \\
\end{array} \right. \]
optimal value for $c = 2.1$, and

\noindent
Tukey's biweight
\[ \Psi(u) = \left\{
\begin{array}{ll}
u\left(1 - \frac{u^2}{c^2}\right)^2, & |u| < c \\
0, & |u| > c
\end{array} \right. \]
optimal value for $c = 5-6$.

The minimization occurs iteratively,
the initial guess for $m$ being the median of the data set,
and for the scale $s$ the normalized median absolute deviation (1.48$\cdot$MAD).
All estimations presented in the paper were obtained with Tukey's biweight.
Other $\Psi$-functions provide similar results.

\end{appendix}

\end{document}